\documentclass[aps,prd,preprintnumbers,groupedaddress,nofootinbib,amssymb,eqsecnum,notitlepage]{revtex4}
\usepackage{here}
\usepackage{graphicx}
\usepackage{amsmath,amsthm,amssymb}
\usepackage{bm}
\usepackage{color}

\usepackage{amsfonts}
\usepackage{dcolumn}
\usepackage{hyperref}
\allowdisplaybreaks[1]
\usepackage{ulem}

\begin{document}
\newcommand{\newc}{\newcommand}

\newcommand{\rk}{\textcolor{red}}
\newc{\ben}{\begin{eqnarray}}
\newc{\een}{\end{eqnarray}}
\newc{\be}{\begin{equation}}
\newc{\ee}{\end{equation}}
\newc{\ba}{\begin{eqnarray}}
\newc{\ea}{\end{eqnarray}}
\newc{\D}{\partial}
\newc{\rH}{{\rm H}}
\newc{\rd}{{\rm d}}

\title{Stability of relativistic stars with scalar hairs}

\author{Ryotaro Kase$^{1}$, Rampei Kimura$^{2}$, 
Seiga Sato$^{3}$, and Shinji Tsujikawa$^{3}$}

\affiliation{
$^1$Department of Physics, Faculty of Science, 
Tokyo University of Science, 1-3, Kagurazaka,
Shinjuku-ku, Tokyo 162-8601, Japan\\
$^2$Waseda Institute for Advanced Study, Waseda University
19th building, 1-21-1 Nishiwaseda, Shinjuku-ku, Tokyo 169-0051, Japan\\
$^3$Department of Physics, Waseda University, 3-4-1 Okubo, Shinjuku, Tokyo 169-8555, Japan}

\begin{abstract}

We study the stability of relativistic stars in scalar-tensor theories with 
a nonminimal coupling of the form $F(\phi)R$, where $F$ 
depends on a scalar field $\phi$ and $R$ is the Ricci scalar. 
On a spherically symmetric and static background, we incorporate a 
perfect fluid minimally coupled to gravity as a form of the Schutz-Sorkin action. 
The odd-parity perturbation for the multipoles $l \geq 2$ is ghost-free under 
the condition $F(\phi)>0$, with the speed of gravity equivalent to that of light. 
For even-parity perturbations with $l \geq 2$, there are three 
propagating degrees of freedom arising from the perfect-fluid, scalar-field, 
and gravity sectors. For $l=0, 1$, the dynamical degrees of freedom 
reduce to two modes. We derive no-ghost conditions and the propagation speeds 
of these perturbations and apply them to concrete theories of hairy relativistic stars 
with $F(\phi)>0$. As long as the perfect fluid satisfies a weak energy 
condition with a positive propagation speed squared $c_m^2$, 
there are neither ghost nor Laplacian instabilities for theories of spontaneous 
scalarization and Brans-Dicke (BD) theories with a BD parameter 
$\omega_{\rm BD}>-3/2$ (including $f(R)$ gravity). 
In these theories, provided $0<c_m^2 \le 1$, we show that all the propagation speeds 
of even-parity perturbations are sub-luminal inside the star, 
while the speeds of gravity outside the star are equivalent to that of light.

\end{abstract}

\date{\today}


\maketitle

\section{Introduction}
\label{introsec}

The dawn of gravitational-wave (GW) astronomies \cite{Abbott2016} 
shed new light on the physics around compact objects such as black holes and neutron stars (NSs). 
For example, the GW170817 event \cite{GW170817} arising from binary NSs placed constraints 
on the mass-radius relation of NSs by their tidal deformations. 
The accumulation of GW events in the future will allow us to probe  
the accuracy of General Relativity (GR) in the strong 
gravitational regime \cite{Berti,Barack}. 
In particular, whether or not some extra degrees of freedom 
are present around compact objects are a great concern, 
along with the problem of dark energy and dark matter.

The simplest candidate for such an extra degree of freedom is 
a scalar field $\phi$. Theories in which the scalar field is coupled 
to the gravitational sector are called scalar-tensor theories \cite{Fujii}. 
The nonminimal coupling $F(\phi)R$, where $F$ is a function of 
$\phi$ and $R$ is the Ricci scalar, is a typical example of 
the direct interaction between the scalar field and gravity. 
In the presence of baryonic fluids, 
the matter sector indirectly feels an interaction with the 
scalar field through the coupling to gravity. 
This matter coupling manifests itself after performing a so-called 
conformal transformation to the Einstein frame in which 
the Ricci scalar does not have a direct coupling 
to $\phi$ \cite{DeFelice:2010aj}. 
In scalar-tensor theories, such matter couplings 
can modify the internal structure of relativistic stars.

In scalar-tensor theories with the nonminimal coupling $F(\phi)R$, 
it is known that there are some NS solutions with scalar hairs 
on a spherically symmetric and static background. 
For the theories in which the coupling $F(\phi)$ contains an 
even power-law function $\phi^n$, 
there exists a field profile of nonvanishing $\phi$, while satisfying 
$F_{,\phi}(0)=0$ as in GR (where $F_{,\phi}={\rm d}F/{\rm d}\phi$) \cite{Damour,Damour2}.
If the second derivative $F_{,\phi \phi}$ is positive at $\phi=0$, then 
the GR branch ($\phi=0$ everywhere) can be unstable due to 
a negative mass squared proportional to $-F_{,\phi \phi}(0)$. 
This allows a possibility for triggering tachyonic growth of the 
scalar field toward a scalarized branch with nonvanishing $\phi$, whose 
phenomenon is dubbed spontaneous scalarization.

A typical example of the nonminimal coupling triggering 
spontaneous scalarization is of the form 
$F(\phi)=e^{-\beta \phi^2/(2M_{\rm pl}^2)}$ \cite{Damour,Damour2}, 
where $\beta$ is a constant and $M_{\rm pl}$ is the reduced Planck mass. 
Provided that $\beta<0$, the two conditions $F_{,\phi}(0)=0$ 
and $F_{,\phi \phi}(0)>0$ are satisfied.
More precisely, spontaneous scalarization can occur for the coupling 
$\beta<-4.35$ \cite{Harada:1998ge,Novak:1998rk,Silva:2014fca,Freire:2012mg}, 
whose upper bound is insensitive to the change of equations of state inside the Ns. 
The properties of hairy solutions and its observational consequences 
were investigated in several contexts, e.g., 
GW asteroseismology \cite{Sotani1,Sotani2}, 
rotating NSs \cite{Sotani:2012eb,Doneva:2013qva,Doneva:2014faa,Pani:2014jra}, 
and the influence on particle geodesics around NSs \cite{Doneva:2014uma,Pappas:2015npa}.
Recent studies showed that black holes can also exhibit 
spontaneous scalarization, in the presence of couplings 
to a Gauss-Bonnet term \cite{Kleihaus:2015aje,Doneva:2017bvd,Silva:2017uqg,Antoniou:2017acq,Antoniou:2017hxj,Minamitsuji:2018xde,Cunha} and
to an electromagnetic field \cite{Stefanov,Herdeiro1,Herdeiro2,Herdeiro3,Ikeda}.

There are also other nonminimally coupled theories admitting hairy NS 
solutions, e.g., Brans-Dicke (BD) theories \cite{Brans} with a scalar potential. 
The nonminimal coupling in 
BD theories can be expressed in the form $F(\phi)=e^{-2Q\phi/M_{\rm pl}}$, 
where $Q$ is a constant related to the so-called BD parameter $\omega_{\rm BD}$ 
as $2Q^2=1/(3+2\omega_{\rm BD})$ \cite{Yoko}. 
Since this coupling does not satisfy conditions for the occurrence of 
spontaneous scalarization, there exist only  
hairy solutions with nonvanishing scalar-field 
profiles \cite{Cooney:2009rr,Arapoglu:2010rz,Orellana:2013gn,Astashenok:2013vza,Yazadjiev:2014cza,Resco:2016upv}. 
For example, $f(R)$ gravity belongs to a class of BD theories with 
$\omega_{\rm BD}=0$ \cite{Ohanlon,Chiba03}, 
in the presence of a scalar potential arising from the deviation from GR.
For the Starobinsky model $f(R)=R+R^2/(6m^2)$ \cite{Staro}, the existence of 
a positive constant mass $m$ gives rise to an exponential growing mode 
of $\phi$ outside the star \cite{Ganguly,Kase:2019dqc}. 
This is not the case for the models $f(R)=R+aR^p$ with 
$1<p<2$ \cite{Kase:2019dqc,Dohi:2020bfs}, 
in which the effective mass of $\phi$ can approach 0 toward spatial infinity.
In the latter case, there are hairy NS solutions with the mass-radius relation 
modified from that in GR.

The hairy NS solutions in theories of spontaneous scalarization and 
BD theories have an additional pressure induced by a matter coupling
with the scalar field. Under the occurrence of spontaneous scalarization, 
for example, the additional  pressure works to be repulsive against gravity, 
in which case the radius and mass of star tend to be increased relative to those 
in GR (see Refs.~\cite{Maselli,Minami,Chagoya,Kase:2017egk,Chagoya2,Ogawa,Kase:2020yhw} 
for related papers). 
Naively, one might expect that such hairy solutions can be unstable 
against perturbations.
To study the stability of relativistic stars with scalar hairs, it is necessary 
to properly incorporate perturbations of the matter sector besides 
those of gravity and the scalar field. 
In this paper, we will address this issue by dealing with baryonic matter
as a perfect fluid.

For a k-essence scalar field with the Lagrangian $P(X)$ \cite{Scherrer}, where $P$ is 
a function of the field kinetic energy $X=-\partial_{\mu}\phi \partial^{\mu}\phi/2$, 
it is known that the corresponding energy-momentum tensor reduces to 
that of a perfect fluid for a time-like scalar field, i.e., $X>0$ \cite{Hu05,Arroja}. 
This is the case for a time-dependent cosmological background, 
so that the k-essence Lagrangian was extensively used to describe the 
perfect-fluid dynamics especially in the context of late-time cosmic 
acceleration \cite{DMT10,KT14b,Heisenberg:2016eld,Gum}. 
On the spherically symmetric and static background, however, the scalar field 
is space-like ($X<0$) and hence the perfect fluid cannot be described by 
the k-essence Lagrangian. Instead, we will employ the matter action advocated by 
Schutz and Sorkin \cite{Sorkin}, which allows one to describe the perfect fluid 
in any curved background (see also Refs.~\cite{Brown,DGS}) .

To accommodate both theories of spontaneous scalarization and BD theories 
presented in Sec.~\ref{theorysec}, we will consider 
scalar-tensor theories given by the Lagrangian ${\cal L}=G_4(\phi)R+G_2(\phi,X)$ 
in the presence of a perfect fluid. 
This belongs to a subclass of Horndeski theories with second-order 
field equations of motion \cite{Horndeski,Horn1,Horn2,Horn3}.
We carry out all the analysis in the Jordan frame, 
in which the matter sector is minimally coupled to gravity. 
After deriving covariant equations of motion and applying them to the 
spherically symmetric and static background in Sec.~\ref{scasec}, we proceed to 
the discussion about the separation of perturbations into odd- and 
even-parity modes in Sec.~\ref{persec}. 
In Horndeski theories without matter, the stabilities of hairy black hole solutions 
against odd- and even-parity perturbations were studied in 
Refs.~\cite{DeFelice:2011ka,Kobayashi:2012kh,Kobayashi:2014wsa,Kobayashi:2014eva,
Ogawa:2015pea,Babichev:2016rlq,Khoury:2020aya}.

In Sec.~\ref{oddsec}, we expand the full action up to second order in 
odd-parity perturbations with the multipoles $l \ge 2$ and show that the 
ghost is absent for $G_4>0$ with the propagation speed of gravity 
equivalent to that of light. 
In Sec.~\ref{evensec}, we derive conditions for the absence of ghosts 
and Laplacian instabilities in the even-parity sector. The propagating degrees of 
freedom are different depending on the multipoles $l$, so we separately discuss 
the three cases $l \ge 2$, $l=1$, and $l=0$. 
The analysis of odd- and even-parity perturbations was also performed 
in Ref.~\cite{Sotani:2005qx} for the nonminimal coupling of Refs.~\cite{Damour,Damour2}
by using an energy-momentum tensor of the perfect fluid. 
Since this approach is based on the equations of motion rather than the action principle,
it is not straightforward to identify the dynamical degrees of freedom and their
stability conditions. In this paper, we address this problem by dealing with the perfect 
fluid as the Schutz-Sorkin action and integrate out all the nondynamical 
perturbations from the action. 
Indeed, the dynamical degree of freedom 
in the matter sector is of a nontrivial form, which affects the propagation 
of scalar GWs.
 
In Sec.~\ref{staconsec}, we apply our general stability conditions to 
hairy relativistic stars present in theories of spontaneous scalarization 
and BD theories. We show that these hairy solutions are stable under the conditions 
$G_4>0$, $\rho+P>0$, and $c_m^2>0$, where $\rho$, $P$, and $c_m^2$ are 
the density, pressure and sound speed squared of matter respectively. 
In particular, as long as $0<c_m^2 \le 1$, all the speeds of propagation 
associated with even-parity perturbations in the gravity sector 
are subluminal inside the star, while they are equivalent to the speed 
of light outside the star. This fact is consistent with the speed 
of gravity constrained from the GW170817 event \cite{GW170817}.

Throughout the paper, we adopt the natural units for which the speed of 
light $c$, the reduced Planck constant $\hbar$, and the
Boltzmann constant $k_B$ are set to unity. 

\section{Theories of scalarized relativistic stars}
\label{theorysec}

In this section, we briefly review several theories of relativistic stars 
with scalar hairs. We consider a scalar field $\phi$ with a nonminimal coupling of 
the form $F(\phi)R$. The scalar field can have a kinetic term of the form 
$\omega(\phi) X$, where $\omega$ is a function of $\phi$ and 
$X=-(1/2) g^{\mu \nu} \partial_{\mu} \phi \partial_{\nu} \phi$ is the kinetic energy 
(with metric tensor $g^{\mu \nu}$).
We also allow for the existence of a field potential $V(\phi)$.
Then, the action of such scalar-tensor theories is given by 
\be
{\cal S}=\int {\rm d}^4 x \sqrt{-g} \left[ \frac{M_{\rm pl}^2}{2} 
F(\phi)R+\omega(\phi)X-V(\phi) \right]
+{\cal S}_m (g_{\mu \nu}, \Psi_m)\,,
\label{Jaction}
\ee
where $g$ is a determinant of metric tensor $g_{\mu \nu}$, and 
$M_{\rm pl}$ is the reduced Planck mass. 
The action ${\cal S}_m$ corresponds to that of the matter field 
$\Psi_m$. We assume that the matter field is minimally coupled 
to gravity in the Jordan frame given by the metric $g_{\mu \nu}$. 
In Sec.~\ref{scasec}, we specify the action ${\cal S}_m$ to be 
of the form of a perfect fluid.

Performing a so-called conformal transformation 
$(g_{\mu \nu})_E=F(\phi) g_{\mu \nu}$ of the metric, 
we obtain the following Einstein-frame action \cite{DeFelice:2010aj},
\be
{\cal S}
=\int {\rm d}^4 x \sqrt{-g_E} \left[ \frac{M_{\rm pl}^2}{2} 
R_E-\frac{1}{2} (g^{\mu \nu})_E 
\partial_{\mu} \varphi \partial_{\nu} \varphi -V_E(\varphi) \right]
+{\cal S}_m \left( F^{-1} (\phi)(g_{\mu \nu})_E, \Psi_m \right)\,,
\label{Eaction}
\ee
where the subscript ``$E$'' represents quantities in the Einstein frame, 
and 
\be
\frac{\rd \varphi}{\rd \phi}=\sqrt{\frac{3}{2} 
\left( \frac{M_{\rm pl} F_{,\phi}}{F} \right)^2
+\frac{\omega}{F}}\,,\qquad
V_E(\varphi)=\frac{V}{F^2}\,.
\ee
In the Einstein frame the field $\varphi$ does not have a direct coupling 
with the Ricci scalar $R_E$, but the matter sector has an interaction 
with the scalar field through the metric $(g_{\mu \nu})_E=F(\phi) g_{\mu \nu}$.

In what follows, we will present two classes of theories which belong to 
the action (\ref{Jaction}).

\subsection{Theories of spontaneous scalarization}

When the nonminimal coupling $F(\phi)$ is present, 
spontaneous scalarization can occur inside  
relativistic stars for a massless scalar field $\phi$. 
The canonical scalar field $\varphi$ in the Einstein frame 
can be chosen to be equivalent to $\phi$. 
Since $\rd \varphi/\rd \phi=1$ in this case, it follows that 
$\omega=[1-3M_{\rm pl}^2 F_{,\phi}^2/(2F^2)]F$.
In the absence of the potential $V(\phi)$, 
the Jordan-frame action (\ref{Jaction}) is expressed 
in the form, 
\be
{\cal S}=\int {\rm d}^4 x \sqrt{-g} \left[ \frac{M_{\rm pl}^2}{2} 
F(\phi)R+\left( 1-\frac{3M_{\rm pl}^2 F_{,\phi}^2}
{2F^2} \right) F(\phi) X \right]
+{\cal S}_m (g_{\mu \nu}, \Psi_m)\,.
\label{Jaction2}
\ee

For the theories (\ref{Jaction2}) with $F_{,\phi}(0)=0$, there is a GR branch 
of spherically symmetric and static NS solutions characterized by $\phi=0$ 
everywhere. About the NS solution in GR, the effective mass squared for small perturbations 
is given by $m_{\rm eff}^2=- (M_{\rm pl}^2/2)[F_{,\phi \phi}(0)/\omega (0)]R$.
As long as the conditions $F_{,\phi \phi}(0)>0$ and $\omega (0)>0$ are satisfied 
with a positive $R$, there is a tachyonic instability of the GR branch.
Then, the spontaneous growth of $\phi$ can occur toward the 
other nontrivial branch with $\phi \neq 0$. The conditions for the occurrence of 
spontaneous scalarization correspond to $F_{,\phi}(0)=0$, 
$F_{,\phi \phi}(0)>0$, and $\omega (0)>0$, so it is 
necessary to have an even power-law dependence of 
$\phi$ for the coupling $F(\phi)$.

The nonminimal coupling chosen by Damour and 
Esposito-Farese \cite{Damour,Damour2} is given by 
\be
F(\phi)=e^{-\beta \phi^2/(2M_{\rm pl}^2)}\,,
\label{Fnon}
\ee
where $\beta$ is a constant. 
In this case, the function $\omega(\phi)$ in front of the kinetic term 
$X$ in Eq.~(\ref{Jaction2}) reads
\be
\omega(\phi)=\left( 
1-\frac{3\beta^2 \phi^2}{2M_{\rm pl}^2}\right)
e^{-\beta \phi^2/(2M_{\rm pl}^2)}\,.
\ee
Then, the conditions $F_{,\phi}(0)=0$, $F_{,\phi \phi}(0)>0$, 
and $\omega (0)>0$ are satisfied for $\beta<0$. 
Hence, for negative $\beta$, the NS can have a nontrivial branch 
with a modified internal structure by the presence of nonminimal 
coupling with the scalar field.

A common procedure for the analysis of spontaneous 
scalarization is to study the solutions in the Einstein frame 
first and then transform back to the Jordan frame to 
compute physical quantities such as the mass and radius of relativistic 
stars \cite{Damour,Damour2,Harada:1998ge,Novak:1998rk,Silva:2014fca,
Freire:2012mg,Chen,Morisaki}. 
However, all the analysis can be performed in the Jordan-frame 
action (\ref{Jaction2}) without any reference to the Einstein frame.
In the Jordan frame the matter sector is minimally coupled to gravity, 
so it is also straightforward to incorporate it as a perfect fluid 
described by a Schutz-Sokin action (see Sec.~\ref{scasec}).

\subsection{Brans-Dicke theories}

The action in BD theories \cite{Brans} with a scalar potential $V(\phi)$ 
can be expressed in the form \cite{Yoko}, 
\be
{\cal S}=\int {\rm d}^4x \sqrt{-g} \left[ \frac{M_{\rm pl}^2}{2} F(\phi)R
+ \left( 1-6Q^2 \right)F(\phi) X -V(\phi) \right]
+{\cal S}_m (g_{\mu \nu}, \Psi_m)\,,
\label{actionBD}
\ee
where the nonminimal coupling is given by 
\be
F(\phi)=e^{-2 Q\phi/M_{\rm pl}}\,.
\label{Fphi}
\ee
The constant $Q$ characterizes the coupling strength 
between the scalar field $\phi$ and gravity, which is related  
to the BD parameter $\omega_{\rm BD}$ as 
\be
2Q^2=\frac{1}{3+2\omega_{\rm BD}}\,.
\label{Qome}
\ee
The action (\ref{actionBD}) belongs to a sub-class of scalar-tensor 
theories (\ref{Jaction}).
The minimally coupled scalar field in GR
corresponds to the limit $Q \to 0$, i.e., 
$\omega_{\rm BD} \to \infty$. 
In the absence of matter the ghost is 
absent for $\omega_{\rm BD}>-3/2$ \cite{Fujii}, which is consistent 
with the positivity on the left-hand-side of Eq.~(\ref{Qome}).

The metric $f(R)$ gravity is accommodated 
by the action (\ref{actionBD}) with the correspondence \cite{DeFelice:2010aj}, 
\be
Q=-\frac{1}{\sqrt{6}}\,,\qquad 
V(\phi)=\frac{M_{\rm pl}^2}{2} \left( FR-f \right)\,,\qquad 
F=\frac{\partial f}{\partial R}=e^{-2 Q\phi/M_{\rm pl}}\,.
\label{fRre}
\ee
In this case, the action (\ref{actionBD}) 
reduces to ${\cal S}=\int {\rm d}^4x \sqrt{-g}\,
(M_{\rm pl}^2/2) f(R)+{\cal S}_m (g_{\mu \nu}, \Psi_m)$, 
with the BD parameter $\omega_{\rm BD}=0$ \cite{Ohanlon,Chiba03}. 
Provided that $f(R)$ contains nonlinear functions of $R$, 
the gravitational sector propagates one scalar degree of freedom 
$\phi$. This scalar field, which is related to $R$ through the last 
relation of Eq.~(\ref{fRre}), has a potential $V(\phi)$ of the 
gravitational origin.

If the potential has a constant mass $m$ like the Starobinsky model 
$f(R)=R+R^2/(6m^2)$, it is difficult to realize a stable field profile 
satisfying the boundary condition $\phi \to 0$ at spatial infinity 
due to the existence of an exponentially growing mode 
outside the star \cite{Ganguly,Kase:2019dqc}.
If the effective mass of $\phi$ approaches 0 toward 
spatial infinity, there exist regular NS solutions without 
the exponential growth of $\phi$. 
An explicit example of the latter is the model $f(R)=R+aR^p$, 
where $a$ and $p$ are constants in the ranges 
$a>0$ and $1<p<2$ \cite{Kase:2019dqc,Dohi:2020bfs}. 
In this case, the scalar potential is approximately given by 
$V(\phi) \propto \phi^{p/(p-1)}$. 
This includes the self-coupling potential $V(\phi) \propto \phi^4$ 
(for $p=4/3$). 

\section{Scalar-tensor theories with matter}
\label{scasec}

In this paper, we focus on scalar-tensor theories given by the action,   
\be
{\cal S}=\int {\rm d}^4 x \sqrt{-g} \left[ 
G_4(\phi)R+G_2(\phi,X) \right]
+{\cal S}_m (g_{\mu \nu}, \Psi_m)\,,
\label{action}
\ee
where $G_4$ is a function of the scalar field $\phi$, and 
$G_2$ depends on both $\phi$ and $X$. 
The action (\ref{action}) accommodates scalar-tensor theories 
with Eq.~(\ref{Jaction}) as a special case.
A perfect fluid minimally coupled to gravity 
can be described by a Schutz-Sorkin action 
of the form \cite{Sorkin,Brown,DGS}
\be
{\cal S}_{m} =  -\int {\rm d}^{4}x \left[
\sqrt{-g}\,\rho(n)
+ J^{\mu} (\partial_{\mu} \ell+{\cal A}_i\partial_{\mu}{\cal B}^i)\right]\,.
\label{SM}
\ee
The matter density $\rho$ depends on the fluid number density $n$ alone. 
The vector field $J^{\mu}$ corresponds to a current, 
while the scalar quantity $\ell$ is a Lagrange multiplier.
In terms of $J^{\mu}$, the number density can be expressed as
\be
n=\sqrt{\frac{g_{\mu \nu}J^{\mu} J^{\nu}}{g}}\,.
\label{defn}
\ee
The fluid four-velocity $u_{\mu}$ is related to $J_{\mu}$, as 
\be
u_{\mu}=\frac{J_{\mu}}{n\sqrt{-g}}\,.
\label{defu}
\ee
{}From Eq.~(\ref{defn}), there is 
the relation $u^{\mu} u_{\mu}=-1$. 
The quantities ${\cal A}_i$ and ${\cal  B}^i$ are spatial vectors 
characterizing intrinsic vector modes. 

\subsection{Covariant equations of motion}

Varying the action (\ref{action}) with respect to $\ell$ and $J^{\mu}$ 
respectively, we obtain
\ba
\partial_{\mu} J^{\mu} &=&0\,,\label{Jcon}\\
\partial_{\mu} \ell &=& 
\rho_{,n} u_{\mu}
-{\cal A}_i\partial_{\mu}{\cal B}^i\,,
\label{rhomu}
\ea
where we used the property $\partial n/\partial J^{\mu}=-u_{\mu}/\sqrt{-g}$.
Here and in the following, we use the notation $\rho_{,n}=\partial \rho/\partial n$.
Variations of the action (\ref{SM}) with respect to ${\cal A}_i$ 
and ${\cal B}^i$ lead, respectively, to 
\ba
&&
J^{\mu}\partial_{\mu}{\cal B}^i=0\,,\label{Bd}\\
&&
J^{\mu}\partial_{\mu}{\cal A}_i=0\,,\label{Ad}
\ea
where we used Eq.~(\ref{Jcon}).
Taking note of the relations $J^{\mu} =n \sqrt{-g}\,u^{\mu}$ 
and $\partial_{\mu}(\sqrt{-g}\,u^{\mu})
=\sqrt{-g}\,\nabla_{\mu}u^{\mu}$, where $\nabla_{\mu}$ is the 
covariant derivative operator, 
the current conservation (\ref{Jcon}) can be expressed 
in the form, 
\be
u^{\mu}\nabla_{\mu}\rho
+(\rho+P)\nabla_{\mu}u^{\mu}=0\,, 
\label{umu}
\ee
where $P$ is the matter pressure defined by 
\be
P \equiv n \rho_{,n}-\rho\,.
\ee

Variation of the matter Lagrangian 
$L_{m}=-[\sqrt{-g}\,\rho(n)
+ J^{\mu} (\partial_{\mu} \ell
+{\cal A}_i\partial_{\mu}{\cal B}^i)]$
with respect to $g^{\mu \nu}$ gives
\be
\delta L_m=-\delta \sqrt{-g} \rho (n)-\sqrt{-g} \rho_{,n} \delta n 
-J_{\mu} \left(  \partial_{\nu} \ell+{\cal A}_i\partial_{\nu}{\cal B}^i
\right) \delta g^{\mu \nu}\,.
\ee
By exploiting Eq.~(\ref{rhomu}) as well as the relations, 
\be
\delta \sqrt{-g} 
= -\frac{1}{2} \sqrt{-g}\,
g_{\mu \nu} \delta g^{\mu \nu}\,,\qquad
\delta n = \frac{n}{2} \left( g_{\mu \nu} 
-u_{\mu} u_{\nu} \right) \delta g^{\mu \nu}\,,
\ee
it follows that 
\be
-\frac{2}{\sqrt{-g}} \frac{\delta L_m}{\delta g^{\mu \nu}}
=\left( \rho+P \right) u_{\mu} u_{\nu}+P g_{\mu \nu}
\equiv T_{\mu \nu}\,,
\label{Tmunu}
\ee
where $T_{\mu \nu}$ corresponds to an energy-momentum 
tensor of the perfect fluid.

Varying the total action (\ref{action}) with respect to $g^{\mu \nu}$, 
the resulting gravitational equations of motion are given by 
\be
2G_4 G_{\mu \nu}-G_2g_{\mu\nu}-G_{2,X}\nabla_{\mu}\phi\nabla_{\nu}\phi
-2G_{4,\phi}\left(\nabla_{\mu}\nabla_{\nu}\phi-g_{\mu\nu}\Box\phi\right)
-2G_{4,\phi\phi}\left(\nabla_{\mu}\phi\nabla_{\nu}\phi
+2Xg_{\mu\nu}\right)=T_{\mu \nu}\,,
\label{Gmu}
\ee
where $\square=g^{\mu \nu} \nabla_{\mu}\nabla_{\nu}$. 
Since the perfect fluid is minimally coupled to gravity, the 
corresponding energy-momentum tensor is 
divergence-free, i.e., 
\be
\nabla^{\mu} T_{\mu \nu}=0\,.
\label{Tcon}
\ee
This property is consistent with the left-hand-side of Eq.~(\ref{Gmu}). 
Multiplying $\mu^{\nu}$ for Eq.~(\ref{Tcon}), 
we obtain
\be
u^{\nu} \nabla^{\mu} T_{\mu \nu}
=-\left[ u_{\mu}\nabla^{\mu}\rho
+(\rho+P)\nabla^{\mu}u_{\mu} \right]=0\,,
\label{Tmueq1}
\ee
which also follows from  Eq.~(\ref{umu}).
Since Eq.~(\ref{umu}) arises from Eq.~(\ref{Jcon}), 
Eq.~(\ref{Tmueq1}) corresponds to the current conservation.

Let us also introduce a unit vector $n^{\nu}$ orthogonal to 
$u_{\nu}$, such that 
\be
n^{\nu}u_{\nu}=0\,,\qquad 
n^{\nu}n_{\nu}=1\,.
\label{ncon}
\ee
Multiplying $n^{\nu}$ for Eq.~(\ref{Tcon}), we find
\be
n^{\nu} \nabla^{\mu} T_{\mu \nu}
=(\rho+P)n^{\nu}u_{\mu} \nabla^{\mu} u_{\nu}
+n_{\mu} \nabla^{\mu} P=0\,,
\ee
and hence
\be
n_{\mu} \nabla^{\mu} P
=-(\rho+P)n^{\nu}u_{\mu} \nabla^{\mu} u_{\nu}\,.
\label{Tmueq2}
\ee
Inside a compact object, this can be interpreted 
as a balance between the pressure and gravity.

\subsection{Background equations}

Let us consider a spherically symmetric and 
static background given by the 
line element, 
\be
\rd s^2=-f(r) \rd t^{2} +h(r)^{-1} \rd r^{2}
+ r^{2} \left(\rd \theta^{2}+\sin^{2}\theta\,\rd\varphi^{2} 
\right)\,,
\label{BGmetric}
\ee
where $f(r)$ and $h(r)$ are functions of $r$.
For the matter sector, we take the following 
configuration,
\be
J^{\mu}=\left( \sqrt{-g}\,N(r),0,0,0 \right)\,, \qquad
{\cal A}_i=0\,,
\label{JBG}
\ee
where $\sqrt{-g}=f^{1/2}h^{-1/2}r^2\sin{\theta}$ and 
$N$ is a function of the radial coordinate $r$.
We note that $J^{\mu}$ is not a four vector since 
the second term on the right-hand-side of Eq.~(\ref{SM}) is not 
multiplied by the volume factor $\sqrt{-g}$. If we alternatively define 
$\tilde{J}^{\mu}=J^{\mu}/\sqrt{-g}$, this can be regarded as 
a four vector whose component depends on $r$ alone, i.e.,  
$\tilde{J}^{\mu}=(N(r),0,0,0)$. 
This is the reason why $J^{\mu}$ contains the $\theta$-dependent 
term arising from $\sqrt{-g}$.

Assuming that $N(r)$ is positive, the number density 
(\ref{defn}) reads
\be
n(r)=f(r)^{1/2}N(r)\,,
\label{nBG}
\ee
which depends on $r$.
Substituting Eqs.~(\ref{JBG}) and (\ref{nBG}) into Eq.~(\ref{defu}), 
it follows that  
\be
u^{\mu}=\left( f(r)^{-1/2},0,0,0 \right)\,.
\ee
{}From the definition (\ref{Tmunu}) of the fluid 
energy-momentum tensor, we have
\be
T^{\mu}_{\nu}={\rm diag} \left( -\rho(r),P(r),P(r),P(r) \right)\,.
\ee

{}From the $tt$, $rr$, $\theta \theta$ components 
Eq.~(\ref{Gmu}), we obtain
\ba
& &
\left( \frac{2G_4}{r}+G_{4,\phi}\phi'  \right)h'+2 \left( G_{4,\phi}\phi''
+\frac{2G_{4,\phi}\phi'}{r}+G_{4,\phi \phi} \phi'^2 \right)h
+\frac{2G_4(h-1)}{r^2}-G_2=-\rho\,,\label{back1}\\
& &
\left( \frac{2G_4}{r}+ G_{4,\phi}\phi' \right)h\frac{f'}{f}
+\frac{4G_{4,\phi}\phi' h}{r}+\frac{2G_4(h-1)}{r^2}
-G_2-G_{2,X}\phi'^2 h=P\,,\label{back2}\\
& &
\frac{G_4 h}{2f^2} \left( 2ff''-f'^2 \right)
+\left( \frac12 G_4 h'+\frac{G_4 h}{r}+G_{4,\phi}\phi' h
\right) \frac{f'}{f}+\left( \frac{G_4}{r}+G_{4,\phi} \phi' \right)h' \nonumber \\
& &
+2 \left( G_{4,\phi \phi} \phi'^2+G_{4,\phi}\phi''+\frac{G_{4,\phi}\phi'}{r} 
\right)h -G_2=P\,,\label{back3}
\ea
where a prime represents the derivative with respect to $r$. 
The $\varphi \varphi$ component of Eq.~(\ref{Gmu}) gives the same 
equation as (\ref{back3}).
We note that Eq.~(\ref{Tmueq1}) is trivially satisfied on the background (\ref{BGmetric}). 
For the unit vector $n_{\nu}$ obeying the property (\ref{ncon}), we can choose
$n_{\mu}=(0,h^{-1/2},0,0)$. Then, Eq.~(\ref{Tmueq2}) reduces to 
\be
P'+\frac{f'}{2f} \left( \rho+P \right)=0\,.
\label{back4}
\ee
The equation of motion for the scalar field follows by varying the action (\ref{action}) 
with respect to $\phi$. This amounts to substituting the $r$ derivative of 
Eq.~(\ref{back2}) as well as Eqs.~(\ref{back1}) and (\ref{back2}) into Eq.~(\ref{back4}), 
with the elimination of $P$ from Eqs.~(\ref{back2}) and (\ref{back3}) to solve for $G_4$. 
Then, the scalar field obeys the following equation, 
\ba
{\cal E}_{\phi} &\equiv& 
G_{2,\phi} + G_{2,X} h \left[ \phi'' + {h'\over 2h}  \phi' + \left({2 \over r} + {f' \over 2f}\right)  \phi'\right]
+  G_{2,\phi X}h \phi'^2
-  G_{2,XX} \left( \phi'' +{h'\over 2h}  \phi'\right)h^2 \phi'^2 \nonumber\\
&&+
G_{4, \phi}\left[{2\over r^2 } (1-h-r h') - \left(h' +{4h \over r}\right) {f'\over 2f} 
+ {f'^2 \over 2f^2} h -{f'' \over f}h \right]
=0\,.
\label{back5}
\ea

If the equation of state, 
\be
P=P(\rho)\,,
\label{EOS}
\ee
is given inside a star, we can integrate Eqs.~(\ref{back1})-(\ref{back4}) with Eq.~(\ref{EOS}) 
to solve for $f$, $h$, $\phi$, $P$, and $\rho$. 
The integration is performed up to the radius of star (characterized by $P=0$) 
with the regular boundary conditions 
$f'=h'=\phi'=P'=\rho'=0$ at $r=0$.

\section{Perturbations on the spherically symmetric and static background}
\label{persec}

To study the stability of relativistic stars around the spherically 
symmetric and static space-time (\ref{BGmetric}), we consider metric perturbations 
$h_{\mu \nu}$ on the background metric $\bar{g}_{\mu\nu}$, such that 
\be
g_{\mu \nu}=\bar{g}_{\mu\nu}+h_{\mu \nu}\,.
\ee
There are also perturbations of the scalar field $\phi$ and quantities 
appearing in the Schutz-Sorkin action (\ref{SM}). 
In terms of  the spherical harmonics $Y_{lm}(\theta, \varphi)$, 
the scalar field can be expressed in the form,
\be
\phi=\bar{\phi}(r)+\sum_{l,m} \delta \phi_{lm} (t,r) 
Y_{lm} (\theta,\varphi)\,,
\label{phidec}
\ee
where $\bar{\phi}(r)$ is the background value, and $\delta \phi_{lm}$ 
is the perturbed part being a function of $t$ and $r$. 
In the following, we omit the subscripts $l, m$ from the perturbation 
$\delta \phi_{lm}(t,r)$ and also apply the same rule to other perturbed quantities 
appearing below.
Under the rotation in two-dimensional plane ($\theta, \varphi$), 
any scalar perturbation has the parity $(-)^l$, which 
is called the even mode \cite{Regge:1957td,Zerilli:1970se}. 
The odd mode corresponds to a perturbation with the parity  $(-)^{l+1}$. 
The metric components $h_{tt}, h_{tr}, h_{rr}$ transform as scalars under 
a two-dimensional rotation in the ($\theta, \varphi$) plane, so they only 
possess even-parity modes.

The $\theta$ and $\varphi$ components of any vector field 
$V_\mu$ contain both even- and odd-modes, whereas 
the temporal and radial components $V_t$ and $V_r$ 
possess the even mode alone.
To accommodate the odd-parity contribution to $V_{\mu}$, 
we introduce the tensor $E_{ab}=\sqrt{\gamma}\,\varepsilon_{ab}$, where 
$\gamma$ is the determinant of two dimensional metric 
$\gamma_{ab}$ (with $a,b$ either $\theta$ or $\varphi$), and 
$\varepsilon_{ab}$ is the anti-symmetric symbol 
with $\varepsilon_{\theta\varphi}=1$. 
The $\theta, \varphi$ components of fluid four velocity $u_{\mu}$, 
for example, can be expressed as
\ba
u_a &=&\nabla_a v+E_{ab}\nabla^b \tilde{v} 
=\sum_{l,m} v(t,r) \nabla_a Y_{lm}(\theta,\varphi)
+\sum_{l,m} \tilde{v}(t,r)E_{ab} \nabla^b Y_{lm}(\theta,\varphi)\,,
\label{uade}
\ea
where, in the second line, we expressed two scalars $v$ and $\tilde{v}$ 
in terms of the expansion of spherical harmonics.
The first and second terms of Eq.~(\ref{uade}) correspond to 
even- and odd-parity perturbations, respectively. 
The metric components 
$h_{t\theta}, h_{t\varphi}, h_{r \theta}, 
h_{r \varphi}$ transform as vectors
under the two-dimensional rotation in the ($\theta, \varphi$) plane, 
so they can be also expressed in terms of the sum of 
even- and odd-modes analogous to Eq.~(\ref{uade}).

The components $T_{ab}$ of any symmetric tensor $T_{\mu \nu}$ contain 
both even- and odd-modes. 
For instance, the metric components $h_{ab}$ are written 
in the form, 
\ba
h_{ab} 
&=& K g_{ab}+\nabla_{a}\nabla_{b}G+\frac{1}{2} \left( {E_a}^c \nabla_c \nabla_b U
+{E_b}^c  \nabla_c \nabla_a U \right)\nonumber\\
&=& 
\sum_{l,m}\left[K(t,r) g_{ab}Y_{lm}+G(t,r)\nabla_a\nabla_bY_{lm} \right]+
\frac{1}{2}\sum_{l,m}
U (t,r)
\left[
E_{a}{}^c \nabla_c\nabla_b Y_{lm}
+ E_{b}{}^c \nabla_c\nabla_a Y_{lm}
\right]\,,
\ea
where, in the second line, we expressed the three scalars $K$, $G$, and $U$ 
in terms of the expansion of spherical harmonics.

In summary, metric perturbations $h_{\mu \nu}$ and 
perturbed quantities present in the action (\ref{action}) can be
decomposed into even- and odd-modes in the following way.

\subsection{Odd-parity perturbations}

The components of odd-mode metric perturbations 
are written as
\ba
& &
h_{tt}=h_{tr}=h_{rr}=0\,,\\
& &
h_{ta}=\sum_{l,m}Q(t,r)E_{ab}
\nabla^bY_{lm}(\theta,\varphi)\,,
\qquad
h_{ra}=\sum_{l,m}W(t,r)E_{ab} \nabla^bY_{lm}(\theta,\varphi)\,,\\
& &
h_{ab}=
\frac{1}{2}\sum_{l,m}
U (t,r) \left[
E_{a}{}^c \nabla_c\nabla_b Y_{lm}(\theta,\varphi)
+ E_{b}{}^c \nabla_c\nabla_a Y_{lm}(\theta,\varphi)
\right]\,.
\ea
The vector field $J_{\mu}$ in the Schutz-Sorkin action (\ref{SM}) 
has the following components associated with the 
odd-parity sector, 
\be
J_t=\bar{J}_t\,,\qquad
J_{r}=0\,,\qquad 
J_{a}=\sum_{l,m} \sqrt{-\bar{g}}\,\delta j(t,r)E_{ab}
\nabla^bY_{lm}(\theta,\varphi)\,,
\label{Jodd}
\ee
where 
\be
\bar{J}_t=-n(r) \sqrt{f(r)} \sqrt{-\bar{g}}\,,
\ee
with the background value 
$\sqrt{-\bar{g}}=\sqrt{f(r)/h(r)} r^2 \sin \theta$.
The intrinsic vectors ${\cal A}_i$ and ${\cal B}^i$ are 
expressed in the form, 
\be
{\cal A}_i=\delta {\cal A}_i\,,\qquad 
{\cal B}^i=x^i+\delta {\cal B}^i\,,
\label{ABi}
\ee
with the odd-parity perturbed components, 
\ba
&&
\delta {\cal A}_r=0\,,\qquad 
\delta {\cal A}_a=\sum_{l,m}\delta {\cal A}(t,r)
E_{ab} \nabla^bY_{lm}(\theta,\varphi)\,,\\
&&
\delta {\cal B}^r=0\,,\qquad 
\delta {\cal B}^a=\sum_{l,m}\delta {\cal B}(t,r)
{E^{a}}_{b} \nabla^bY_{lm}(\theta,\varphi)\,,
\ea
where the term $x^i$ in ${\cal B}^i$ is normalized 
such that the background contribution to 
$\partial_{j}{\cal B}^i$ reduces to $\delta^i_{j}$.

Let us consider the infinitesimal gauge transformation
$x_\mu\to x_\mu+\xi_\mu$, where
\be
\xi_t=\xi_r=0,
\qquad
\xi_a=\sum_{l,m}\Lambda(t,r) E_{ab} 
\nabla^b Y_{lm}(\theta,\varphi)\,.
\label{xivec}
\ee
Then, the metric perturbations $Q$, $W$, and $U$
transform, respectively, as
\be
Q \to Q+\dot{\Lambda}\,,\qquad
W \to W+\Lambda'
-\frac{2}{r}\Lambda\,,\qquad
U \to U+2\Lambda\,,
\label{gaugetra}
\ee
where a dot represents a derivative with respect to $t$.
For the multipoles $l \geq 2$, we choose the 
Regge-Wheeler gauge \cite{Regge:1957td} characterized by 
\be
U=0\,.
\label{U0}
\ee
For the dipole ($l=1$), the perturbation $h_{ab}$ vanishes identically, 
so we need to handle this case separately.

\subsection{Even-parity perturbations}

The components of metric perturbations in the even-parity sector 
are given by 
\ba
&&
h_{tt}=f(r) \sum_{l,m}H_0(t,r)Y_{lm}(\theta,\varphi)\,,\qquad
h_{tr}=h_{rt}=\sum_{l,m}H_1(t,r)Y_{lm}(\theta,\varphi)\,,\qquad
h_{rr}=h(r)^{-1}\,\sum_{l,m}H_2(t,r)Y_{lm}(\theta,\varphi)\,,\notag\\
&&
h_{ta}=h_{at}=\sum_{l,m}\beta(t,r)\nabla_aY_{lm}(\theta,\varphi)\,,\qquad
h_{ra}=h_{ar}=\sum_{l,m}\alpha(t,r)\nabla_aY_{lm}(\theta,\varphi)\,,\notag\\
&&
h_{ab}=\sum_{l,m}\left[K(t,r)g_{ab}Y_{lm}(\theta,\varphi)
+G(t,r)\nabla_a\nabla_bY_{lm}(\theta,\varphi)\right]\,.
\label{meteven}
\ea
The scalar field $\phi$ is expressed in the form (\ref{phidec}). 
The components of $J_{\mu}$ containing even-parity perturbations
are of the forms, 
\be
J_{t}=\bar{J}_t+\sum_{l,m} \sqrt{-\bar{g}}\,\delta J_t (t,r)Y_{lm}(\theta,\varphi)\,,\qquad
J_{r}=\sum_{l,m} \sqrt{-\bar{g}}\,\delta J_r (t,r)Y_{lm}(\theta,\varphi)\,,\qquad
J_{a}=\sum_{l,m} \sqrt{-\bar{g}}\,\delta J(t,r) \nabla_aY_{lm}(\theta,\varphi)\,.
\label{Jeven}
\ee
The intrinsic spatial vector fields ${\cal A}_i$ and ${\cal B}_i$ 
are given by Eq.~(\ref{ABi}) with the perturbed components,
\ba
& &
\delta {\cal A}_r=\sum_{l,m}\delta {\cal A}_1(t,r)Y_{lm}(\theta,\varphi)\,,\qquad 
\delta {\cal A}_a=\sum_{l,m}\delta {\cal A}_2(t,r)\nabla_aY_{lm}(\theta,\varphi)\,,
\label{delAa}\\
&&
\delta {\cal B}_r=\sum_{l,m}\delta {\cal B}_1(t,r)Y_{lm}(\theta,\varphi)\,,\qquad 
\delta {\cal B}_a=\sum_{l,m}\delta {\cal B}_2(t,r)\nabla_aY_{lm}(\theta,\varphi)\,.
\label{delBa}
\ea

Let us consider the infinitesimal gauge transformation
$x_\mu\to x_\mu+\xi_\mu$, where 
\be
\xi_t=\sum_{l,m} {\cal T}(t,r)Y_{lm}(\theta,\varphi)\,,\qquad 
\xi_r=\sum_{l,m} {\cal R}(t,r)Y_{lm}(\theta,\varphi)\,,\qquad 
\xi_a=\sum_{l,m} \Theta(t,r) \nabla_a Y_{lm}(\theta,\varphi)\,.
\label{xivec2}
\ee
Then, the perturbations $H_0$, $H_1$, $H_2$, 
$\beta$, $\alpha$, $K$, $G$, and 
$\delta \phi$ transform, respectively, 
as \cite{Kobayashi:2014wsa,Motohashi:2011pw}
\ba
& &
H_0 \to H_0+\frac{2}{f} \dot{{\cal T}}-\frac{f' h}{f}{\cal R}\,,\qquad 
H_1 \to H_1+\dot{{\cal R}}+{\cal T}'-\frac{f'}{f}{\cal T}\,,\qquad
H_2 \to H_2+2h{\cal R}'+h' {\cal R}\,,\label{H1tra}\\
& &
\beta \to \beta+{\cal T}+\dot{\Theta}
-\frac{f'}{f}{\cal T}\,,\qquad 
\alpha \to \alpha+{\cal R}+\Theta'
-\frac{2}{r}\Theta\,,\qquad
K \to K+\frac{2}{r}h{\cal R}\,,\qquad 
G \to G+\frac{2}{r^2}\Theta\,,\label{beal} \\
& &
\delta \phi \to \delta \phi-\bar{\phi}' h{\cal R}\,,
\ea
where a dot represents the derivative with respect to $t$. 
For the multipoles $l \geq 2$, the transformation scalars 
${\cal T}$ and $\Theta$ can be 
fixed by choosing the gauge:
\be
\beta=0\,,\qquad G=0\,.
\ee
To fix the other transformation scalar ${\cal R}$, 
there are several different gauge 
choices listed below.
\ba
& &
{\rm (i)~Uniform~curvature~gauge}:~K=0\,,\\
& &
{\rm (ii)~Unitary~gauge}:~\delta \phi=0\,,\\
& &
{\rm (iii)~Spatially~diagonal~gauge}:~\alpha=0\,.
\ea
The physics is not affected by different choices of gauges.
As in Refs.~\cite{DeFelice:2011ka,Motohashi:2011pw,Kobayashi:2014wsa},  
we will choose the uniform curvature gauge (i) 
to compute the second-order action of even-parity perturbations.

The above argument of gauge fixings is valid for the multipoles $l \geq 2$. 
For the monopole ($l=0$), the perturbations $\beta$, $\alpha$, and $G$ vanish identically.
For the dipole ($l=1$), the perturbations in $h_{ab}$ appear only as the combination 
$K-G$, so the decomposition into the two components $K$ and $G$ is redundant. 
We will separately study these cases in Sec.~\ref{evensec}.

\subsection{Matter density perturbation and velocity potential}

We discuss the structure of perturbations in the Schutz-Sorkin action 
in more detail. 
The matter density perturbation $\delta \rho$ is related to 
the perturbation $\delta n$ of fluid number density given 
by Eq.~(\ref{defn}). 
The components (\ref{Jodd}) of $J_{\mu}$ in the odd-parity sector 
do not give rise to the first-order perturbation of $n$. 
On the other hand, the components (\ref{Jeven}) in the even-parity 
sector generate the first-order perturbation of $n$.
From this first-order perturbation $\delta n$, 
we define the matter density perturbation 
$\delta \rho (t,r)$, as 
\be
\delta n=\sum_{l,m} \frac{\delta \rho (t,r)}{\rho_{,n}(r)} 
Y_{lm} (\theta, \varphi)\,.
\label{delnrho}
\ee
After the expansion of $n$ in terms of even-mode perturbations, 
we will convert $\delta n$ to $\delta \rho (t,r)$ for computing the 
second-order action.

For the quantity $\ell$ appearing in the action (\ref{SM}), 
we will derive its explicit form by using the constraint (\ref{rhomu}).
Up to first order in perturbations, the partial derivatives of $\ell$ 
with respect to $a=\theta, \varphi$ are given by 
\be
\partial_a \ell=\rho_{,n} u_a-\delta {\cal A}_a\,,
\label{pal}
\ee
where we used Eq.~(\ref{ABi}). 
Since we are considering the derivative of the scalar quantity $\ell$, 
we only need to consider the even-mode contribution 
to $u_a$, i.e., the first term in Eq.~(\ref{uade}).
On using the second of Eq.~(\ref{delAa}) and integrating Eq.~(\ref{pal}) 
with respect to $a$, it follows that 
\be
\ell={\cal F}(t,r)+\sum_{l,m} \left[ \rho_{,n}(r) v(t,r)-\delta {\cal A}_2 
(t,r) \right] Y_{lm} (\theta, \varphi)\,,
\label{elle}
\ee
where ${\cal F}$ is a function of $t$ and $r$. 
The term $\rho_{,n}(r) v(t,r)$ in Eq.~(\ref{elle}) corresponds to the first-order 
quantity, so that $\rho_{,n}$ should be evaluated on the background. 
The time derivative $\partial_t \ell$ is equivalent to 
$\rho_{,n} (r)\bar{u}_t(r)=-\rho_{,n} (r) \sqrt{f(r)}$ at the background level.
The integration of the relation $\partial_t {\cal F}=-\rho_{,n} (r) \sqrt{f(r)}$ 
gives ${\cal F}(t,r)=-\rho_{,n} (r) \sqrt{f(r)}\,t$, so that 
\be
\ell=-\rho_{,n} (r) \sqrt{f(r)}\,t+\sum_{l,m} \left[ \rho_{,n}(r) v(t,r)-\delta {\cal A}_2 
(t,r) \right] Y_{lm} (\theta, \varphi)\,.
\label{ldef}
\ee
The temporal and radial components of four velocity $u_{\mu}$ 
can be expressed, respectively, as
\be
u_t=-\sqrt{f(r)}+\sum_{l,m}\delta u_t (t, r) 
Y_{lm} (\theta, \varphi)\,,\qquad 
u_r=\sum_{l,m}\delta u_r (t, r) Y_{lm} (\theta, \varphi)\,,
\ee
where $\delta u_t $ and $\delta u_r$ are functions of $t$ and $r$.
Since $\partial_{t}\ell=\rho_{,n}u_{t}$ up to first order in perturbations, 
we obtain the correspondence,
\be
\delta u_t=\sqrt{f(r)} c_m^2 
\frac{\delta \rho}{\rho+P}+\dot{v}\,,
\label{ut}
\ee
where $c_m^2$ is the matter sound speed squared defined by 
\be
c_m^2 \equiv \frac{n \rho_{,nn}}{\rho_{,n}}\,.
\label{cm}
\ee
Similarly, we take the $r$ derivative of Eq.~(\ref{ldef}) 
and compare it with the relation $\partial_{r}\ell=\rho_{,n}u_r-\delta{\cal A}_r$. 
In doing so, we exploit the property,
\be
\frac{\rd}{\rd r}
\left (\rho_{,n}(r) \sqrt{f(r)} \right)
=\sqrt{f(r)} \left(\rho_{,n}'+\frac{f'}{2f} \rho_{,n} \right)=0\,,
\ee
where the second equality follows from Eq.~(\ref{back4}) 
with $\rho_{,n}'=P'/n$. 
Then, the perturbation $\delta u_r$ can be expressed as 
\be
\delta u_r=v'-\frac{f'}{2f} v
+\frac{\delta {\cal A}_1-\delta {\cal A}_2'}{\rho_{,n}}\,.
\label{ur}
\ee
Equations (\ref{ut}) and (\ref{ur}) show 
the correspondence between the perturbations $\delta \rho$, 
$\delta {\cal A}_1$, $\delta {\cal A}_2$ and the components 
of four velocity $u_{\mu}$.
Since $u_{\mu}=J_{\mu}/(n\sqrt{-g})$, there are also relations
between the perturbations $\delta J_r$, $\delta J$ 
in Eq.~(\ref{Jeven}) and $v$, $\delta {\cal A}_1$, 
$\delta {\cal A}_2$ appeared above.
In Sec.~\ref{evensec}, we will address this issue.

\section{Odd-parity perturbations}
\label{oddsec}

We expand the action (\ref{action}) up to second-order in odd-parity perturbations. 
In doing so, we choose the Regge-Wheeler gauge (\ref{U0}) for $l \geq 2$. 
For the dipole ($l=1$) the condition $U=0$ automatically holds, 
so we study this case separately at the end of this section.

The scalar field $\phi$ does not possess the odd-parity perturbation, so we can 
use the background value of $G_4(\phi)$ for the expansion of $G_4(\phi)R$. 
The field kinetic energy $X$ is expanded as
\be
X=-\frac12 h \phi'^2-\frac{h^2}{2r^2} \phi'^2 W^2
\left[ (\partial_{\theta}Y_{lm})^2+
\frac{(\partial_{\varphi}Y_{lm})^2}{\sin^2\theta}\right]
+{\cal O}(\varepsilon^4)\,,
\label{Xexpan}
\ee
where $\varepsilon^n$ represents the $n$-th order of perturbations. 
We expand the k-essence Lagrangian as 
$G_2(\phi,X)=G_2(r)+G_{2,X}\delta X$, where 
$\delta X$ is the second-order perturbation in Eq.~(\ref{Xexpan}).

The fluid number density (\ref{defn}) can be decomposed into the 
background part $n(r)$ and the second-order perturbed part 
$\delta n$ given by 
\be
\delta n=\frac{n}{2r^2}\left(hW^2-\frac{2}{f}Q^2
+\frac{2}{n\sqrt{f}}Q\delta j-\frac{\delta j^2}{n^2}\right)
\left[ (\partial_{\theta}Y_{lm})^2+\frac{(\partial_{\varphi}
Y_{lm})^2}{\sin^2\theta} \right]
+{\cal O}(\varepsilon^4)\,.
\ee
Then, the matter density $\rho$ in the action (\ref{SM}) 
contains the second-order perturbation $\rho_{,n}(r) \delta n$.
As we derived in Eq.~(\ref{ldef}), the quantity $\ell$ 
has only even-mode perturbations and hence 
$\ell=-\rho_{,n} (r) \sqrt{f(r)}\,t$ for odd modes. 
We caution that the vector component $J^t$ contains the 
second-order odd-parity perturbation $\delta J^t$ besides the background 
value $\bar{J}^t=n(r)r^2 \sin \theta/\sqrt{h}$. 
The perturbations $\delta J^{\theta}$ and  $\delta J^{\varphi}$ 
correspond to the first-order perturbation.
Hence the terms $\delta J^t \dot{\ell}+
\sum_{a=\theta,\varphi} (\bar{J}^t \delta {\cal A}_a \dot{\delta {\cal B}^a}
+\delta J^a \delta {\cal A}_a)$ give rise to the second-order 
contribution to Eq.~(\ref{SM}).

After expanding Eq.~(\ref{action}) up to quadratic order 
in odd-parity perturbations, the resulting second-order action 
contains terms multiplied by $\delta {\cal A}(t,r)$.
Varying this action with respect to $\delta {\cal A}$, it follows that 
\be
\dot{\delta {\cal B}}=\frac{nQ-\sqrt{f}\,\delta j}{nr^2}\,.
\ee
After substituting this relation into the second-order action, the 
terms related to $\delta j$ appear as the quadratic dependence $\delta j^2$.
Hence the variation of the action with respect to $\delta j$ gives 
\be
\delta j=0\,.
\ee
In this way, the perturbations $\delta {\cal A}$, $\delta {\cal B}$, $\delta j$ 
are integrated out from the second-order action. 

The next step is to perform the integral with respect to $\theta$ and $\varphi$. 
In this procedure, it is sufficient to set $m=0$ and multiply the action by 
$2\pi$. As for the integration with respect to $\theta$, 
we use the formulas of integrals of $Y_{l0}$ and 
their $\theta$ derivatives given in Appendix B of Ref.~\cite{Kase:2018voo}. 
The resulting quadratic-order action contains the $t$ and $r$ derivatives 
of perturbations $W$ and $Q$, so we integrate some of them by parts. 
By using the background Eqs.~(\ref{back1}) and (\ref{back2}), 
the second-order action of odd-parity perturbations reduces to
\be
{\cal S}^{(2)}_{\rm odd}=\sum_{l,m}
\int \rd t \rd r \left[ \frac{L}{2} \sqrt{\frac{h}{f}}G_4
\left( \dot{W}-Q'+\frac{2Q}{r} \right)^2
-\frac{L(L-2)\sqrt{fh}\,G_4}{2r^2}W^2
+\frac{L(L-2)G_4}{2\sqrt{fh}\,r^2}Q^2 \right]\,,
\label{Sodd}
\ee
where
\be
L=l(l+1)\,.
\ee
{}From Eq.~(\ref{Sodd}), we observe that the presence of the perfect fluid 
does not affect the evolution of odd-mode perturbations.
In the following, we will study the two different cases:
(i) $l \geq 2$ and (ii) $l=1$, in turn.

\subsection{$l \geq 2$}

The variation of Eq.~(\ref{Sodd}) with respect to the nondynamical 
variable $Q$ leads to a constraint equation for $Q$.
However, the presence of the $Q'^2$ term does not allow one 
to solve explicitly for $Q$. 
To overcome this problem, we introduce the Lagrange multiplier 
$\chi(t,r)$ and express the action (\ref{Sodd}) in the form, 
\be
{\cal S}^{(2)}_{\rm odd}=\sum_{l,m} \int 
\rd t \rd r \left\{ \frac{L}{2} \sqrt{\frac{h}{f}}G_4
\left[ 2\chi \left( \dot{W}-Q'+\frac{2Q}{r} \right)-\chi^2 
\right]
-\frac{L(L-2)\sqrt{fh}\,G_4}{2r^2}W^2
+\frac{L(L-2)G_4}{2\sqrt{fh}\,r^2}Q^2 \right\}\,.
\label{Sodd2}
\ee
Varying the action (\ref{Sodd2}) with respect to $W$ and $Q$, 
respectively, we obtain
\ba
W &=& -\frac{r^2}{f(L-2)} \dot{\chi}\,,\\ 
Q &=& -\frac{[(h' \chi+2h \chi')r+4h\chi]G_4 f
-f'h r \chi G_4+2fh r \phi' \chi G_{4,\phi}}{2(L-2)fG_4} r\,.
\ea
Substituting these relations and their $t$ and $r$ derivatives into 
Eq.~(\ref{Sodd2}) and integrating it by parts, the second-order 
action is expressed in the form, 
\be
{\cal S}^{(2)}_{\rm odd}=\sum_{l,m} \int 
\rd t \rd r \left( {\cal K} \dot{\chi}^2+{\cal G} \chi'^2
+{\cal M} \chi^2 \right)\,,
\label{Sodd3}
\ee
where
\ba
{\cal K} &=& \frac{L\sqrt{h}r^2 G_4}{2f^{3/2} (L-2)}\,,\qquad
{\cal G} = -\frac{Lh^{3/2}r^2 G_4}{2\sqrt{f} (L-2)}\,,\\
{\cal M} &=& -L\sqrt{h} [\{ (2L+4h-4-h'' r^2-4h'r)f^2
+(f''h+f'h')fr^2-f'^2h r^2 \}G_4^2 \nonumber \\
& &
-2\{ (G_{4,\phi \phi} \phi'^2+G_{4,\phi}\phi'')h
+G_{4,\phi}h' \phi'\} f^2 r^2G_4
+2f^2 h r^2 \phi'^2 G_{4,\phi}^2]/[4f^{5/2}(L-2)G_4]\,.
\ea
There is one propagating degree of freedom $\chi$ arising from the 
gravitational sector. 
Since $L \geq 6$ for $l \geq 2$,  
the condition for the absence of ghosts corresponds to ${\cal K}>0$, i.e., 
\be
{\cal F} \equiv 2G_4>0\,.
\label{nogoodd}
\ee

Let us derive the propagation speed of $\chi$ along the radial direction.
Assuming the solution of the form $\chi=e^{i (\omega t -kr)}$ 
and taking the limits of large $\omega$ and $k$, 
the dispersion relation following from Eq.~(\ref{Sodd3}) reads
\be
\omega^2 {\cal K}+k^2{\cal G}=0\,.
\ee
The propagation speed squared in terms of the coordinates $t$ and $r$ 
is $\hat{c}_r^2=\omega^2/k^2=-{\cal G}/{\cal K}=fh$.
The propagation speed $c_r$ along the radial direction in proper time 
$\tau=\int \sqrt{f} \rd t$ is given by $c_r=\rd r_*/\rd \tau$, 
where $\rd r_*=\rd r/\sqrt{h}$. 
Since $c_r$ is related to $\hat{c}_r=\rd r/\rd t$ as $c_r=\hat{c}_r/\sqrt{fh}$, 
it follows that 
\be
c_r^2=1\,.
\ee
Hence there is no Laplacian instability of the odd-mode perturbation $\chi$ 
in the radial direction.

Along the angular direction, we employ the solution of the form 
$\chi=e^{i(\omega t-l \theta)}$.
For large multipoles ($l \gg 1$), the term ${\cal M}\chi^2$ in Eq.~(\ref{Sodd3}) 
contributes to the dispersion relation besides the term ${\cal K} \dot{\chi}^2$, so that 
\be
\omega^2 {\cal K}+{\cal M}=0\,,
\ee
where 
\be
{\cal K} = \frac{\sqrt{h}r^2 G_4}{2f^{3/2}}+ {\cal O}(l^{-2})\,,\qquad
{\cal M}=-\frac{\sqrt{h} G_4 }{2\sqrt{f}}\,l^2+ {\cal O}(l)\,.
\ee
In terms of the time coordinate $t$, the propagation speed squared along the 
angular direction is $\hat{c}_{\Omega}^2=\omega^2 r^2/l^2=
-({\cal M}/{\cal K}) r^2/l^2=f$, where 
we have taken the limit $l \to \infty$ in the last equality.
In proper time, the propagation speed is given by 
$c_{\Omega}=r \rd \theta/\rd \tau=\hat{c}_{\Omega}/\sqrt{f}$. 
Hence, in the limit $l \gg 1$, we obtain 
\be
c_{\Omega}^2=1\,.
\ee
This means that there is no Laplacian instability along the 
angular direction either. 
We have thus shown that the stability of the odd-mode perturbation 
$\chi$ is ensured under the no-ghost condition $G_4>0$. 
For the theories (\ref{action}), the speed of odd-parity GWs is equivalent to that of light.

\subsection{$l=1$}

For the dipole there is the relation $U=0$, so we have a residual gauge 
degree of freedom. Since $L=2$ in this case, the last two terms in 
the square bracket of Eq.~(\ref{Sodd}) vanish. 
Then, the second-order action reduces to 
\be
{\cal S}^{(2)}_{\rm odd}=\sum_{l,m} \int 
\rd t \rd r \sqrt{\frac{h}{f}}G_4
\left( \dot{W}-Q'+\frac{2Q}{r} \right)^2\,.
\label{Sodd4}
\ee
For the gauge choice $W=0$,  the transformation 
scalar $\Lambda$ in Eq.~(\ref{gaugetra}) is given by 
\be
\Lambda (t,r)=-r^2 \int {\rm d} \tilde{r} \frac{W (t, \tilde{r})}
{\tilde{r}^2}+r^2 {\cal C}_1(t)\,,
\label{Lamin}
\ee
where ${\cal C}_1(t)$ is an arbitrary function of $t$ 
corresponding to a gauge mode.
We vary the action (\ref{Sodd4}) with respect to $W$ and $Q$, 
and set $W=0$ in the end. This leads to 
\ba
\dot{\cal U} &=& 0\,,\label{calU1}\\
\left( r^2 {\cal U} \right)' &=& 0\,,\label{calU2}
\ea
where
\be
{\cal U} (t,r)\equiv \sqrt{\frac{h}{f}}G_4
\left( Q'-\frac{2Q}{r} \right)\,.
\label{Utr}
\ee
The integrated solutions to Eqs.~(\ref{calU1}) and (\ref{calU2}) are 
given by 
\be
{\cal U}=\frac{C}{r^2}\,,
\label{Uso}
\ee
where $C$ is a constant. 
Substituting Eq.~(\ref{Uso}) into Eq.~(\ref{Utr}) and solving it for $Q$, 
it follows that 
\be
Q(t,r)=r^2 \int \rd \tilde{r} \frac{C}{\tilde{r}^4}
\sqrt{\frac{f}{h}}\frac{1}{G_4}+r^2{\cal C}_2(t)\,,
\label{Qin}
\ee
where ${\cal C}_2(t)$ is an arbitrary function of $t$ corresponding to the 
gauge mode. 
The gauge modes appearing in Eqs.~(\ref{Lamin}) and (\ref{Qin}) 
can be eliminated by choosing 
\be
\dot{\cal C}_1(t)={\cal C}_2(t)\,.
\ee
Substituting Eq.~(\ref{Uso}) into Eq.~(\ref{Sodd4}) with the gauge choice $W=0$, 
we obtain  
\be
{\cal S}^{(2)}_{\rm odd}=\sum_{l,m} \int 
\rd t \rd r \sqrt{\frac{f}{h}}
\frac{C^2}{r^4 G_4}\,,
\label{Sodd5}
\ee
which means that there is no dynamical propagating degree of freedom 
for $l=1$.

We note that the perturbation $\dot{W}-Q'+2Q/r$ appearing 
in the action (\ref{Sodd4}) is gauge-invariant.
In terms of the field $\chi$ introduced in Eq.~(\ref{Sodd2}), the variation of 
the action with respect to $W$ gives $\dot{\chi}=0$. 
Since $\chi$ depends on $r$ alone, the gauge-invariant 
perturbation $\chi=\dot{W}-Q'+2Q/r$ does not work 
as a dynamical perturbation. 
This is consistent with the argument given above.

\section{Even-parity perturbations}
\label{evensec}

We proceed to the derivation of stability conditions in the even-parity sector
by expanding the action up to second order in perturbations.
Since the second-order action is different depending 
on the multipoles $l$, we will discuss the three cases: 
(A) $l \geq 2$, (B) $l=0$, and (C) $l=1$, in turn.

\subsection{$l \geq 2$}
\label{secA}

For $l  \geq 2$, we choose the uniform curvature gauge given by 
\be
K=0\,,\qquad \beta=0\,,\qquad G=0\,,
\ee
under which ${\cal T}$, ${\cal R}$, and $\Theta$ in the gauge transformation 
(\ref{beal}) are fixed. In the gravity sector, we are left with four metric perturbations 
$H_0$, $H_1$, $H_2$, and $\alpha$. 
For the perfect fluid, we consider the vector field $J_{\mu}$ in the form 
(\ref{Jeven}) and adopt the configuration (\ref{ABi}) with the perturbations 
$\delta {\cal A}_i$ and $\delta {\cal B}_i$ given by 
Eqs.~(\ref{delAa}) and (\ref{delBa}). 
{}From Eq.~(\ref{ldef}), the Lagrange multiplier $\ell$ contains the 
velocity potential $v(t,r)$ and the perturbation 
$\delta {\cal A}_2 (t,r)$. This expression of $\ell$ is used 
for expanding the Schutz-Sorkin action.
The matter perturbation $\delta \rho$ is related to the perturbation of 
number density $\delta n$, as Eq.~(\ref{delnrho}). 
The density $\rho$ in the Schutz-Sorkin action is expanded 
in the form, 
\be
\rho=\rho(r)+\rho_{,n}\delta n
+\frac{\rho_{,n}}{2n}c_m^2  \delta n^2+
{\cal O}(\varepsilon^3)\,,
\ee
where $c_m^2$ is defined by Eq.~(\ref{cm}). 

In the scalar-field sector, we perform the expansions, 
\ba
G_2(\phi,X) &=& 
G_2(r)+G_{2,\phi}\delta \phi+G_{2,X}\delta X
+\frac{1}{2} G_{2,\phi \phi} \delta \phi^2
+\frac{1}{2} G_{2,X X} \delta X^2
+G_{2,\phi X} \delta \phi \delta X+{\cal O}(\varepsilon^3)\,,\\
G_4(\phi)
&=&
G_4(r)+G_{4,\phi}\delta \phi+\frac{1}{2} G_{4,\phi \phi} 
\delta \phi^2+{\cal O}(\varepsilon^3)\,,
\ea
where $\delta \phi=\sum_{l,m} \delta \phi (t, r)Y_{lm}$, and
\ba
\delta X &=&
\sum_{l,m} \frac12 h \phi' \left( \phi' H_2- 2\delta \phi' 
\right) Y_{lm}
+\sum_{l,m} \frac{1}{2f} \left[ \dot{\delta \phi}^2+h^2 \phi'^2 H_1^2
-h \left\{ f\delta \phi'^2+2\phi' (H_1 \dot{\delta \phi}-f H_2 \delta \phi' )
+f \phi'^2 H_2^2 \right\} \right] Y_{lm}^2 \nonumber\\
&&-\sum_{l,m}\frac{1}{2r^2} \left( \delta \phi-h \phi' \alpha \right)^2 
\left[ (\partial_{\theta}Y_{lm})^2+\frac{(\partial_{\varphi}
Y_{lm})^2}{\sin^2\theta} \right]
+{\cal O}(\varepsilon^3)\,.
\ea
Since it is sufficient to consider the mode $m=0$, we will do 
so in the following discussion.
We expand the total action (\ref{action}) up to second order in 
even-mode perturbations. 
Varying the resulting second-order action with respect to 
$\delta {\cal B}_1$, $\delta {\cal B}_2$, 
$\delta J$, and $\delta J_r$, respectively, it follows that  
\ba
& &
\dot{\delta {\cal A}}_1=0\,,
\label{percon2}\\
& &
\dot{\delta {\cal A}}_2=0\,,
\label{percon3}\\
& &
n v-\delta J=0\,,
\label{percon4}\\
& &
2 \left( \rho+P \right) \delta J_r-2n^2 \left( \delta {\cal A}_1
-\delta {\cal A}_2' \right)
-n  \left( \rho+P \right) \left( 2v' -\frac{f'}{f} v \right)=0\,.
\label{percon5}
\ea
We solve Eqs.~(\ref{percon4}) and (\ref{percon5}) for 
$\delta J$ and $\delta J_r$, respectively, and 
substitute them into the total second-order action. 
After this procedure, there exist the terms proportional to 
$\delta {\cal A}_1 \dot{\delta {\cal B}}_1$ and 
$\delta {\cal A}_2 \dot{\delta {\cal B}}_2$ with time-independent coefficients, 
but they can be integrated out on account of Eqs.~(\ref{percon2}) and (\ref{percon3}). 
The second-order action containing the perturbations $\delta {\cal A}_1$ 
and $\delta {\cal A}_2$ reduces to 
${\cal S}_{\delta {\cal A}_{1,2}}=
\sum_{l,m} \int \rd t \rd r {\cal L}_{\delta {\cal A}_{1,2}}$, where
\be
{\cal L}_{\delta {\cal A}_{1,2}}
=\frac{n r^2}{2 (\rho+P)} \sqrt{\frac{h}{f}}
\left[ (\rho+P) \left( 2\sqrt{f} H_1+f' v-2f v' \right)
(\delta {\cal A}_1-\delta {\cal A}_2') 
-nf \left( \delta {\cal A}_1-\delta {\cal A}_2' \right)^2 \right]\,.
\label{SdelA}
\ee
The combination $\delta {\cal A}_1-\delta {\cal A}_2'$ can be replaced 
with the perturbation $\delta u_r$ given by Eq.~(\ref{ur}). 
Then, varying the action (\ref{SdelA}) with respect to $\delta u_r$, 
we obtain
\be
\delta u_r=\frac{H_1}{\sqrt{f}}\,.
\ee
On using this relation, the Lagrangian (\ref{SdelA}) can be expressed 
in terms of $v$, its $r$ derivative, and $H_1$, as
\be
{\cal L}_{\delta {\cal A}_{1,2}}
=\frac{(\rho+P)r^2 \sqrt{h}}{8 f^{3/2}}
\left( 2 \sqrt{f} H_1 +f'v-2 fv'\right) ^2 \,.
\label{SdelA2}
\ee
In the full quadratic-order action, there are also terms containing the perturbations 
$v$ and $\delta \rho$, which arise from  
$J^\mu \partial_{\mu} \ell$ and $\rho(n)$ in Eq.~(\ref{SM}).

On using the background Eqs.~(\ref{back1})-(\ref{back3}) and performing 
the integration by parts, the second-order action of even-mode 
perturbations can be expressed in the form 
${\cal S}_{\rm even}^{(2)}=\sum_{l,m} \int \rd t \rd r {\cal L}$, where
\ba
{\cal L}
&=& H_0 \left[ a_1 \delta \phi'' +a_2 \delta \phi' +a_3 H_2'
+L a_4 \alpha'+\left( a_5+L a_6 \right) \delta \phi 
+\left( a_7+L a_8 \right) H_2+L a_9 \alpha +a_{10} \delta \rho \right] \nonumber \\
& &+L b_1 H_1^2+H_1 \left( b_2 \dot{\delta \phi}'+b_3 \dot{\delta \phi}
+b_4 \dot{H}_2+L b_5 \dot{\alpha} \right)
+c_1 \dot{\delta \phi} \dot{H}_2
+H_2 \left[ c_2 \delta \phi'+ (c_3+L c_4) \delta \phi
+L c_5 \alpha+\tilde{c}_5 \dot{v} \right]+c_6 H_2^2 \nonumber \\
& &+L d_1 \dot{\alpha}^2+L \alpha \left( d_2 \delta \phi'
+d_3 \delta \phi \right)+L d_4 \alpha^2
+e_1 \dot{\delta \phi}^2+e_2 \delta \phi'^2
+\left( e_3+L e_4 \right) \delta \phi^2
+L f_1 v^2+f_2 \delta \rho^2+f_3 \delta \rho\,\dot{v}\,.
\label{eaction}
\ea
The background-dependent coefficients $a_1, a_2, ...$ are explicitly given 
in Appendix.
In comparison to the paper by 
Kobayashi, Motohashi, Suyama (KMS) \cite{Kobayashi:2014wsa} 
without the perfect fluid, 
there is the notational difference of a factor $1/2$, i.e., 
$a_1=a_1^{\rm KMS}/2,...,e_4=e_4^{\rm KMS}/2$, due to the 
different normalization of $Y_{lm}$. 
Each second equality in Eq.~(\ref{evencof}) among 
coefficients (e.g., $a_5=a_2'-a_1''$) is valid even in full Horndeski theories 
containing the dependence of $G_3(\phi,X)$, $G_4(\phi,X)$, 
and $G_5(\phi,X)$ \cite{Kobayashi:2014wsa}.

When the perfect fluid is absent, there are two propagating 
degrees of freedom in the even-parity sector. 
One of them is the field perturbation $\delta \phi$, and 
the other is the following 
combination \cite{DeFelice:2011ka,Kobayashi:2014wsa}, 
\be
\psi \equiv a_3 H_2+L a_4 \alpha+a_1 \delta \phi'\,,
\label{psi}
\ee
which corresponds to the dynamical perturbation in 
the gravity sector. 
The variable $\psi$ is analogous to the dynamical perturbation 
taken by Moncrief \cite{Moncrief} and Zerilli \cite{Zerilli:1970se} in the Regge-Wheeler gauge 
($\alpha=\beta=G=0$). 
While the Moncrief-Zerilli variable \cite{Lousto:1996sx} is the combination of 
$H_2$ and $K$, the perturbation (\ref{psi}) contains 
$H_2$ and $\alpha$. Due to the existence of the term 
$a_1 \delta \phi'$ in Eq.~(\ref{psi}), the 
derivatives $a_1 \delta \phi''$ and $a_3 H_2'$ in Eq.~(\ref{eaction}) 
can be simultaneously replaced with $\psi'$ \cite{Kobayashi:2014wsa}.

In the perfect-fluid sector, we introduce the following 
dynamical matter perturbation, 
\be
\delta \rho_m \equiv \delta \rho+\frac{2\sqrt{f}\,hr^3 f_1}
{f_3[ha_3(2f-f' r)+Lfr a_4]}\psi'
-\frac{h f_1 r^2[a_1(2f-f' r)+f^2 rb_3]}
{\sqrt{f} f_3[ha_3(2f-f' r)+Lfr a_4]}\delta \phi'\,.
\label{delrhom}
\ee
If we try to obtain the second-order action of dynamical perturbations 
in terms of $\delta \rho$, this gives rise to the apparent dynamical terms 
$\dot{\psi}'^2$ and $\dot{\delta \phi}'^2$. 
However,  they can be eliminated by introducing 
the second and third terms on the 
right-hand-side of Eq.~(\ref{delrhom}).

Varying the Lagrangian (\ref{eaction}) with respect to the nondynamical 
perturbations $H_0$, $H_1$, and $v$, respectively, we obtain
\ba
& &
a_1 \delta \phi'' +a_2 \delta \phi' +a_3 H_2'
+L a_4 \alpha'+\left( a_5+L a_6 \right) \delta \phi 
+\left( a_7+L a_8 \right) H_2+L a_9 \alpha +a_{10} \delta \rho=0\,,
\label{econ1}\\
& &
H_1=-\frac{1}{2L b_1} \left( b_2 \dot{\delta \phi}'+b_3 \dot{\delta \phi}
+b_4 \dot{H}_2+L b_5 \dot{\alpha} \right)\,,
\label{econ2}\\
& &
v=\frac{1}{2L f_1} 
\left( f_3 \dot{\delta \rho}+\tilde{c}_5 \dot{H}_2 \right)\,.
\label{econ3}
\ea
Taking the $r$ derivative of Eq.~(\ref{psi}) and substituting it into Eq.~(\ref{econ1}), 
the nondynamical perturbation $\alpha$ can be expressed in terms of 
$\psi$, $\delta \phi$, and $\delta \rho_m$ and their first radial derivatives. 
{}From Eq.~(\ref{psi}), the variable $H_2$ and its 
time derivative are written in terms of $\psi$, $\alpha$, 
$\delta \phi'$ and their first time derivatives. 
We plug these relations and Eqs.~(\ref{econ2})-(\ref{econ3}) into Eq.~(\ref{eaction}).
After the integration by parts, the resulting second-order action is expressed 
in the form,
\be
{\cal S}_{\rm even}^{(2)}=\sum_{l,m} \int \rd t \rd r
\left( 
\dot{\vec{\mathcal{X}}}^{t}{\bm K}\dot{\vec{\mathcal{X}}}
+\vec{\mathcal{X}}^{'t}{\bm G}\vec{\mathcal{X}}^{'}
+\vec{\mathcal{X}}^{t}{\bm Q}\vec{\mathcal{X}}^{'}
+\vec{\mathcal{X}}^{t}{\bm M} \vec{\mathcal{X}}
\right)\,,
\label{Ss2}
\ee
where ${\bm K}$, ${\bm G}$, ${\bm Q}$, ${\bm M}$ 
are $3 \times 3$ matrices, with 
\be
\vec{\mathcal{X}}^{t}=\left( \delta \rho_m, \psi, \delta \phi \right) \,.
\label{calX}
\ee
To derive no-ghost conditions, 
we only resort to relations among the coefficients presented in 
the second equalities of Eq.~(\ref{evencof}) in the Appendix and 
define the following quantities,
\be
\mu \equiv -\frac{4a_3}{\sqrt{fh}}\,,\qquad
{\cal H} \equiv \frac{2a_4}{\sqrt{fh}}\,,\qquad 
{\cal P}_1 \equiv \frac{h \mu}{2fr^2 {\cal H}^2} \left( 
\frac{fr^4 {\cal H}^4}{\mu^2 h} \right)'\,,\qquad
{\cal P}_2 \equiv -h \left( 2-\frac{rf'}{f} \right)\mu\,, 
\ee
where ${\cal H}$, ${\cal P}_1$, and ${\cal P}_2$ are the same 
as those introduced in Ref.~\cite{Kobayashi:2014wsa}.
It is convenient to notice the following relation,
\be
{\cal P}_2'=\frac{r(r{\cal H}'+{\cal H}){\cal H} {\cal P}_2
-2r\mu {\cal H}{\cal F}-\mu {\cal P}_1 {\cal P}_2}{r^2{\cal H}^2}
-\frac{4\mu r }{{\cal H}} \sqrt{\frac{h}{f}}f_1\,,
\ee
where we recall that ${\cal F}=2G_4$ for the theories (\ref{action}).

The ghost is absent under the following three conditions, 
\be
K_{11}>0\,,\qquad K_{11}K_{22}-K_{12}K_{21}>0\,,\qquad 
{\rm det}\,{\bm K}>0\,.
\ee
The first condition corresponds to
\be
K_{11}=\frac{(2r{\cal H}L+{\cal P}_2)^2 r^4}{2\sqrt{fh}\,(\rho+P)L
[2r{\cal H}L+{\cal P}_2-2(\rho+P)r^3]^2}>0\,,
\label{nogo1}
\ee
which is satisfied for $\rho+P>0$.
The second translates to 
\be
K_{11}K_{22}-K_{12}K_{21}=
\frac{4(L{\cal P}_1-{\cal F})\mu^2 r^4}
{f^2{\cal H}^2 L^2 (\rho+P)[2r{\cal H}L+{\cal P}_2-2(\rho+P)r^3]^2}
>0\,.\label{nogo2}
\ee
Provided that $\rho+P>0$, the condition (\ref{nogo2}) holds 
for $L{\cal P}_1-{\cal F}>0$.
Finally, the third condition is given by 
\be
{\rm det}\,{\bm K}=\frac{2r^4[h^2 \mu^2(L-2){\cal F}
(2{\cal P}_1-{\cal F})-r^2 (\rho+P)
\{ {\cal H}^2r^4 L (\rho+P)+2{\cal H}{\cal F} hr \mu (L-2)
-h \mu^2 (L{\cal P}_1-{\cal F})\}]}
{(fh)^{5/2}(\rho+P)L^2{\cal H}^2 \phi'^2
[2r{\cal H}L+{\cal P}_2-2(\rho+P)r^3]^2}
>0\,.\label{nogo3}
\ee
As long as $\rho+P>0$, the condition (\ref{nogo3})
is satisfied for a positive numerator. 
When the perfect fluid is absent, the no-ghost condition 
corresponds to ${\cal F}(2{\cal P}_1-{\cal F})>0$ \cite{Kobayashi:2014wsa}. 
Indeed, this condition can be recovered by taking the limit 
$\rho+P \to +0$ in Eq.~(\ref{nogo3}). 
Adding the perfect fluid modifies the third no-ghost condition.
In the above derivation of no-ghost conditions, we only used the relations 
among coefficients in the second-order action (\ref{eaction}), so 
the results (\ref{nogo1})-(\ref{nogo3})
are valid even in full Horndeski theories with more general coefficients 
$a_1$ etc given in Ref.~\cite{Kobayashi:2014wsa}.

In the limit of large wave number $k$, the three propagation speeds $c_r$ 
along the radial direction in proper time can be obtained by solving 
\be
{\rm det} \left| fh c_r^2 {\bm K}+{\bm G} \right|=0\,.
\label{crso}
\ee
The matrix components $G_{11}$, $G_{12}$, and $G_{13}$ of 
symmetric matrix ${\bm G}$, which are 
related to the matter perturbation $\delta \rho_m$, vanish identically. 
This is attributed to the fact that the velocity potential $v$
in Eq.~(\ref{uade}) arises from the $\theta$ and $\varphi$ 
components of the four velocity $u_{\mu}$.
There is no propagation of the matter perturbation in the radial 
direction, so the corresponding value of $c_{r}^2$ yields
\be
c_{r1}^2=0\,.
\ee
As for the other two radial speeds of propagation, the derivation of 
their general expressions applicable to full Horndeski theories 
is not straightforward due to a mixture of the gravitational propagation 
speed with the perfect-fluid sector. 
Hence we focus on scalar-tensor theories given by the action 
(\ref{action}) in the following.
Then, the two propagation speed squares read
\be
c_{r\pm}^2=\frac{C_2}{2C_1} \left[ -1 \pm 
\sqrt{1-\frac{4C_1 C_3}{C_2^2}} \right]\,,
\label{crgee}
\ee
where
\ba
C_1 &=&
G_{2,X} G_4 \left[4(L-2)G_4 h + L r^2 (\rho + P)\right] + 2 G_{4,\phi}^2 \left[6(L-2)G_4 h + (2L-1)r^2(\rho+P)\right] \,,\\
C_2 &=&
-G_{2,X} G_4 \left[8(L-2)G_4 h +L r^2 (1+c_m^2) (\rho+P) \right]
-2 G_{4,\phi}^2 \left[12 (L-2) G_4 h + r^2 (2L-1) (1+c_m^2) (\rho+P)\right] \notag\\
&&+G_{2,XX} G_4 h \phi'^2 \left[4(L-2)G_4 h + L r^2 (\rho+P)\right] \,,\\
C_3 &=&
G_4 \left[4(L-2)G_4 h + c_m^2 L r^2 (\rho +P)\right](G_{2,X}-G_{2,XX}h \phi'^2) 
+ G_{4,\phi}^2 \left[12 (L-2) G_4 h  + 2(2L-1) c_m^2 r^2 (\rho+P)\right] \,. \notag\\
\ea
If we consider the theories containing only a linear function of 
$X$ in $G_2$, i.e., 
\be
G_{2,XX}=0\,,
\label{G2XX}
\ee
then the propagation speed 
squares (\ref{crgee}) reduce to
\ba
c_{r2}^2 &=& 1-\frac{(1-c_m^2) r^2 [ (L-2)G_{4,\phi}^2 + L(G_{2,X}G_4  + 3 G_{4,\phi}^2)]
(\rho + P)}{4(L-2) G_4 h(G_{2,X}G_4  + 3 G_{4,\phi}^2)+
 r^2 [(L-2)G_{4,\phi}^2 + L(G_{2,X}G_4  + 3 G_{4,\phi}^2)](\rho + P)}\,,
\label{crl>2}\\
c_{r3}^2 &=& 1\,.
\label{crl>2b}
\ea
In theories containing nonlinear functions of $X$ in $G_2$, 
there is the deviation of $c_{r3}^2$ from 1 analogous to 
that of k-essence scalar in Minkowski space-time. 
Then, $c_{r3}^2$ corresponds to the speed of propagation for 
$\delta \phi$, whereas $c_{r2}^2$ to that for $\psi$.
The results (\ref{crl>2}) and (\ref{crl>2b}) are valid for 
scalar-tensor theories given by the action (\ref{Jaction}).
For $\rho+P>0$ and $c_m^2 \neq 1$, there is the deviation of 
$c_{r2}^2$ from 1. 
This property is different from that in Horndeski theories without 
the perfect fluid, in which case the speed of even-parity gravitational 
perturbation $\psi$ is the same  
as that of the odd-parity sector \cite{Kobayashi:2014wsa}. 

The propagation speed $c_{\Omega}$ in the angular direction is
known by solving
\be
{\rm det} \left| l^2fc_{\Omega}^2 {\bm K}
+r^2 {\bm M} \right|=0\,.
\ee
The mass matrix ${\bm M}$ is important only for large $l$, 
so we will take the limit $L=l(l+1) \gg 1$ in the following discussion.
For the theories given by the action (\ref{action}), the propagation 
speed squares $c_{\Omega1}^2$, $c_{\Omega2}^2$, and $c_{\Omega3}^2$ 
of the perturbations $\delta \rho_m$, $\psi$, and $\delta \phi$ are given, respectively, by 
\ba
c_{\Omega1}^2 &=& c_m^2\,,\label{cOl>2a}\\
c_{\Omega2}^2 &=& 1-\frac{r^2 (G_{2,X}G_4+4G_{4,\phi}^2)
(\rho+P)}{4h G_{2,X}G_4^2+[G_{2,X}r^2 (\rho+P)
+12h G_{4,\phi}^2]G_4+4G_{4,\phi}^2 r^2(\rho+P)}\,,\label{cOl>2}\\
c_{\Omega3}^2 &=& 1\,.\label{cOl>2b}
\ea
The speed of propagation in the gravity sector is 
affected by the perfect fluid. 

{}From the above discussions, 
the Laplacian stabilities of even-mode perturbations along the radial and 
angular directions are absent under the conditions 
$c_{r\pm}^2 \geq 0$ and $c_{\Omega2}^2 \geq 0$ 
with $c_m^2 \geq 0$.

\subsection{$l=0$}

For the monopole mode $l=0$, the perturbations $\beta$, $\alpha$, and $G$ 
vanish identically.
We choose the gauge $K=0$ to fix the radial transformation scalar ${\cal R}$ 
in $\xi_r$. 
The second-order Lagrangian for $l=0$ can be derived by setting $L=0$ and $\alpha=0$ 
in Eq.~(\ref{eaction}), such that 
\ba
{\cal L} &=&H_0 \left[ \left( a_1 \delta \phi'-\frac{f}{2}b_3 \delta \phi
-\frac{f}{2}b_4 H_2 \right)'-\frac{r^2}{2}f_1 H_2+a_{10} \delta \rho \right]
+\frac{b_2}{a_1}H_1 \left( a_1 \delta \phi'-\frac{f}{2}b_3 \delta \phi
-\frac{f}{2}b_4 H_2 \right)^{\cdot} \nonumber \\
&&+
c_1 \dot{\delta \phi} \dot{H}_2
+H_2 \left( c_2 \delta \phi'+ c_3 \delta \phi
+\tilde{c}_5 \dot{v} \right)+c_6 H_2^2
+e_1 \dot{\delta \phi}^2+e_2 \delta \phi'^2
+e_3 \delta \phi^2
+f_2 \delta \rho^2+f_3 \delta \rho\,\dot{v}\,.
\label{actionl=0}
\ea
In the following we choose the gauge $H_0=0$. In this case, 
there is a gauge mode ${\cal C}(r)$ in the temporal transformation scalar ${\cal T}$. 
This appears as the gauge mode ${\cal C}'-(f'/f){\cal C}$ in the gauge 
transformation of $H_1$, see Eq.~(\ref{H1tra}).
Varying the action (\ref{actionl=0}) 
with respect to $H_1$, we have
\be
a_1 \delta \phi'-\frac{f}{2}b_3 \delta \phi
-\frac{f}{2}b_4 H_2={\cal D}(r)\,,
\label{const2}
\ee
where ${\cal D}(r)$ is an arbitrary function of $r$. 
The gauge mode ${\cal C}'-(f'/f){\cal C}$ in $H_1$ can be eliminated by properly 
choosing the $r$-dependent function ${\cal D}(r)$. 
This $r$-dependent function depends on the background alone, 
so it does not affect the dynamics of perturbations \cite{Motohashi:2011pw,Kobayashi:2014wsa}. 
Hence we drop such contributions to the second-order action of 
perturbations in the following.
We solve Eq.~(\ref{const2}) for $H_2$ and take the time derivative of $H_2$. 
Substituting $H_2$ and $\dot{H}_2$ into Eq.~(\ref{actionl=0}) and integrating it by 
parts, the action contains the two dynamical fields, 
\be
\vec{\mathcal{X}}^{t}=\left( v,\delta \phi \right)\,.
\ee
After the integration by parts, the reduced action is expressed in the form, 
\be
{\cal S}_{\rm even}^{(2)}=\sum_{l,m} \int \rd t \rd r \left(
\dot{\vec{\mathcal{X}}}^{t}{\bm K}\dot{\vec{\mathcal{X}}}
+\vec{\mathcal{X}}'^{t}{\bm G}\vec{\mathcal{X}}'
+\vec{\mathcal{X}}^{t}{\bm M}\vec{\mathcal{X}}
+\dot{\vec{\mathcal{X}}}^{t}{\bm R}\vec{\mathcal{X}}'
+\dot{\vec{\mathcal{X}}}^{t}{\bm T}\vec{\mathcal{X}} 
\right)\,,
\label{Lre}
\ee
where the nonvanishing components of $2 \times 2$ matrices 
${\bm K}$, ${\bm G}$, ${\bm M}$, ${\bm R}$, and ${\bm T}$ are
\ba
K_{11} &=& \frac{(\rho+P)r^2}{2\sqrt{fh}\,c_m^2}\,,\\
K_{22} &=& \frac{1}{\sqrt{fh} \phi'^2} 
\left[ 2{\cal P}_1-{\cal F}+\frac{r^2 (\rho+P)(\mu-4r{\cal H})}{2h \mu}
\right]\,,\\
G_{22} &=& \frac{a_3^2e_2 -a_1 a_3c_2+a_1^2 c_6}{a_3^2}\,,\\
M_{22} &=& \frac{a_3^2e_3+(a_1'-a_2)[(a_1'-a_2)c_6+a_3 c_3]}{a_3^2}
+\left[ \frac{a_1 a_3 c_3+(a_1'-a_2)(2a_1 c_6-a_3 c_2)}{2a_3^2} \right]'\,,\\
R_{12} &=& R_{21}= -\frac{f_1 a_1 r^2}{2\sqrt{f}\,a_3}\,,\\
T_{12} &=& -T_{21}
=-\frac{r}{8f^{3/2}a_3^2} \left[ a_1(2ff_1r a_3'-4a_ 3 f f_1
-2a_3 f rf_1'+a_3f' r f_1)+2a_3 f f_1 r (2a_2-3a_1')
\right]\,.
\ea
We note that ${\bm R}$ and ${\bm T}$ are symmetric and 
anti-symmetric matrices, respectively.

The ghosts are absent under the two conditions, 
\ba
& &
K_{11}>0\,,\label{nogo1l=0}\\
& &
K_{22}>0\,.\label{nogi2l=0}
\ea
where the former is satisfied for $\rho+P>0$ and $c_m^2>0$. 
On using  $\vec{\mathcal{X}}^{t}=e^{i(\omega t-kr)}$ 
as a solution in the radial direction, 
the dispersion relation for large $\omega$ and $k$ yields
\be
{\rm det}(\omega^2{\bm K}-\omega k {\bm R}+k^2 {\bm G})=0\,.
\label{KRGre}
\ee
The propagation speed $c_r$ in proper time can be obtained by 
substituting $\omega=\sqrt{fh}c_r k$ into Eq.~(\ref{KRGre}). 
The resulting two solutions are given by 
\ba
& &
\left( c_{r1}^2 \right)_{l=0}=0\,,\label{crl=0a}\\
& &
\left(  c_{r2}^2 \right)_{l=0}=-\frac{1}{fh K_{22}} 
\left( G_{22}-\frac{R_{12}^2}{K_{11}} \right)
=1-\frac{4h^2 \phi'^2 G_{2,XX}G_4^2+(1-c_m^2)(\rho+P)r^2 G_{4,\phi}^2}
{4hG_{2,X}G_4^2+[12h G_4+(\rho+P)r^2]G_{4,\phi}^2}\,.
\label{crl=0}
\ea
There is no radial propagation in the perfect-fluid sector, 
but the scalar perturbation $\delta \phi$ propagates with the 
speed $\left( c_{r2} \right)_{l=0}$. 
The Laplacian instability can be avoided for $\left( c_{r2}^2\right)_{l=0} \geq 0$.
Unless $l \gg 1$, 
the kinetic term $\dot{\vec{\mathcal{X}}}^{t}{\bm K}\dot{\vec{\mathcal{X}}}$ dominates 
over the mass term $\vec{\mathcal{X}}^{t}{\bm M}\vec{\mathcal{X}}$ for large $\omega$, 
so we do not consider the propagation along the angular direction.

\subsection{$l=1$}

For the dipole mode $l=1$, the dependence of metric perturbations $h_{ab}$ occurs 
through the combination $K-G$. After fixing the gauge to be $G=0$ 
and $\beta=0$, we can set $K=0$. 
Since the latter does not correspond to the gauge fixing, we will 
choose the gauge $\delta \phi=0$ to fix ${\cal R}$. 
Then, we can simply set $\delta \phi=0$ and $L=2$ in the second-order 
action (\ref{eaction}) and define the dynamical variables, 
\be
\psi \equiv a_3 H_2+L a_4 \alpha\,,\qquad
\delta \rho_m \equiv \delta \rho+\frac{2\sqrt{f}\,hr^3 f_1}
{f_3[ha_3(2f-f' r)+Lfr a_4]}\psi'\,.
\label{psidef2}
\ee
We follow the similar procedure to that taken in 
Eqs.~(\ref{econ1})-(\ref{econ3}) and eliminate the nondynamical variables 
$H_0$, $\alpha$, $H_1$, and $v$. 
After the integration by parts, the resulting second-order action 
is of the form (\ref{Ss2}) with the 
two dynamical perturbations, 
\be
\vec{\mathcal{X}}^{t}=\left( \delta \rho_m, \psi \right) \,.
\label{calX2}
\ee
The no-ghost conditions, which are determined by the 
$2 \times 2$ matrix ${\bm K}$, are   
\ba
K_{11} &=& \frac{(4r{\cal H}+{\cal P}_2)^2 r^4}
{4\sqrt{fh}\,(\rho+P)[4r{\cal H}+{\cal P}_2-2(\rho+P)r^3]^2}>0\,,
\label{nogo1l=1}\\
{\rm det}\,{\bm K} &=&
\frac{(2{\cal P}_1-{\cal F})\mu^2 r^4}
{f^2{\cal H}^2 (\rho+P)[4r{\cal H}+{\cal P}_2-2(\rho+P)r^3]^2}
>0\,.
\label{nogo2l=1}
\ea
Taking the limit $L \to 2$, these results coincide with Eqs.~(\ref{nogo1}) 
and (\ref{nogo2}), respectively.

The propagation speeds along the radial direction 
are known by solving Eq.~(\ref{crso}) for the 
$2 \times 2$ matrices ${\bm K}$ and ${\bm G}$. 
Since the nonvanishing component of ${\bm G}$ is $G_{11}$ alone, 
the propagation speed squared associated with $\delta \rho_m$ is 
\be
\left( c_{r1}^2 \right)_{l=1}=0\,.
\ee
The other solution, which corresponds to the propagation of $\psi$, 
is given by
\be
\left( c_{r2}^2 \right)_{l=1}=\frac{2r^2[\sqrt{f}\,r^2  {\cal H}^2 c_m^2(\rho+P)
-\sqrt{h}(4{\cal H}\mu c_5+8{\cal H}^2 c_6+\mu^2 d_4)]}
{\sqrt{f}h(2{\cal P}_1-{\cal F})\mu^2}\,.
\label{crf}
\ee
For the theories given by the action (\ref{action}), the Laplacian 
instability is absent under the condition, 
\be
\left( c_{r2}^2  \right)_{l=1}=1-\frac{G_4[(1-c_m^2)(\rho+P) 
+h^2 \phi'^4 G_{2,XX}]}{G_4(\rho+P+h \phi'^2 G_{2,X})
+3h \phi'^2 G_{4, \phi}^2} \geq 0\,.
\label{crl=1}
\ee
This is not recovered by taking the limit $L \to 2$ in 
Eq.~(\ref{crgee}), so it gives the additional stability 
condition to that for $l \geq 2$. 

\section{Stability of relativistic stars in concrete theories}
\label{staconsec}

We study the stability of relativistic stars in scalar-tensor theories by using 
the results derived in Secs.~\ref{oddsec} and \ref{evensec}. 
In doing so, we first summarize conditions for the absence of ghosts 
and Laplacian instabilities in the general theories 
(\ref{action}) and apply them to specific theories discussed 
in Sec.~\ref{theorysec}.

First of all, the ghost in the odd-parity sector is absent under the condition 
(\ref{nogoodd}), i.e., 
\be
G_4>0\,.
\label{nog1}
\ee
For even-mode perturbations,  the no-ghost conditions (\ref{nogo1}), (\ref{nogo1l=0}), 
and (\ref{nogo1l=1}), which correspond to stabilities in the matter sector  
for the modes $l \geq 2$, $l=0$, and $l=1$ respectively, are satisfied for
\be
\rho+P>0\,,\qquad c_m^2>0\,.
\label{nog2}
\ee
In the following, we will consider relativistic stars 
composed by baryonic matter obeying 
the inequalities (\ref{nog2}).
The other no-ghost condition for $l=1$, i.e., Eq.~(\ref{nogo2l=1}), 
is the special case of Eq.~(\ref{nogo2}), so we do not need to consider 
the former. Then, the remaining no-ghost conditions are given by 
Eqs.~(\ref{nogo2}), (\ref{nogo3}), and (\ref{nogi2l=0}). 
Under the inequalities (\ref{nog1}) and (\ref{nog2}), 
they translate, respectively, to
\ba
{\cal K}_1 &\equiv& 
(L-2) h (2G_4 +G_{4,\phi} r \phi')^2
+ L r^2 [G_4 (\rho + P ) + \kappa h \phi'^2 ]
>0\,,\label{nog3}\\
{\cal K}_2 &\equiv& 
(L-2)[4 \kappa G_4 h  + G_{4,\phi}^2 r^2 (\rho + P)] + L \kappa r^2  (\rho + P)
>0\,,
\label{nog4}\\
{\cal K}_3 &\equiv& 
4 \kappa G_4 h + G_{4,\phi}^2 r^2 (\rho + P)
>0\,,
\label{nog5}
\ea
where $L>2$ in Eqs.~(\ref{nog3}) and (\ref{nog4}), and 
we defined 
\be
\kappa \equiv
G_{2,X}G_4  + 3 G_{4,\phi}^2 \,.
\label{def:kappa}
\ee

For $l \geq 2$, the propagation speeds of odd-mode perturbations are equivalent to 1. 
The stabilities of even-mode perturbations are ensured 
as long as the speeds of propagation given in Eqs.~(\ref{crl>2}), 
(\ref{cOl>2}), (\ref{crl=0}), and (\ref{crl=1}) are nonnegative. 
In the case with $G_{2,XX}=0$, these conditions translate to
\ba
c_{r2}^2 &=& 
1-\frac{(1-c_m^2) r^2 [ (L-2)G_{4,\phi}^2 + L\kappa](\rho + P) }
{4(L-2) G_4 h \kappa +  r^2 [(L-2)G_{4,\phi}^2 + L\kappa](\rho + P)} \nonumber \\
&=&
\frac{4(L-2) G_4 h \kappa + c_m^2 r^2 [(L-2)G_{4,\phi}^2 + L\kappa](\rho + P) }
{4(L-2) G_4 h \kappa +  r^2 [(L-2)G_{4,\phi}^2 + L\kappa](\rho + P)}
\geq 0 \,,\label{crf1}\\
c_{\Omega 2}^2 &=&
1-\frac{r^2 (G_{4,\phi}^2+\kappa)(\rho+P)}{4G_4 h \kappa + r^2 (G_{4,\phi}^2 + \kappa)(\rho+P)} 
=
\frac{4G_4 h \kappa}{4G_4 h \kappa + r^2 (G_{4,\phi}^2 + \kappa)(\rho+P)} 
\geq 0\,,\\
(c_{r2}^2)_{l=0} &=& 
1-\frac{(1-c_m^2)G_{4,\phi}^2 r^2 (\rho+P)}{4G_4 h \kappa + G_{4,\phi}^2 r^2 (\rho+P)} 
=
\frac{4G_4 h \kappa+c_m^2 G_{4,\phi}^2 r^2 (\rho+P)}
{4G_4 h \kappa + G_{4,\phi}^2 r^2 (\rho+P)}
 \geq 0\,,\\
(c_{r2}^2)_{l=1} &=&
1-\frac{(1-c_m^2) G_4 (\rho + P)}{G_4 (\rho+P) 
+ h \kappa \phi'^2}
=\frac{c_m^2 G_4 (\rho + P) + h \kappa \phi'^2}
{G_4 (\rho+P) +h \kappa \phi'^2}
\geq 0\,.
\label{crf4}
\ea
Now, we discuss stability conditions in two concrete theories
presented in Sec.~\ref{theorysec}.

\subsection{Theories of spontaneous scalarization}

The action in theories of spontaneous scalarization
is given by Eq.~(\ref{Jaction2}), i.e.,  
\be
G_4(\phi)=\frac{M_{\rm pl}^2}{2} F(\phi)\,,\qquad 
G_2(\phi,X)=
\left( 1-\frac{3M_{\rm pl}^2 F_{,\phi}^2}
{2F^2} \right) F(\phi) X\,.
\ee
In this case, the quantity $\kappa$ reduces to
\be
\kappa=\frac{M_{\rm pl}^2}{2} F^2(\phi)>0\,.
\label{kappaSS}
\ee
The positivity of $\kappa$ means that, along with 
the conditions (\ref{nog1}) 
and \eqref{nog2}, the absence of 
ghost and Laplacian instabilities is manifestly guaranteed. 
This is the case for the coupling (\ref{Fnon}) chosen by 
Damour and Esposito-Farese, where $G_4$ is positive.
In addition, as long as $c_m^2$ is in the range 
$0<c_m^2 \le 1$, all the propagation speed 
squares computed above are subluminal.
Since all the conditions \eqref{nog3}-\eqref{nog5} and \eqref{crf1}-\eqref{crf4} 
are irrelevant to the potential, a massive scalar field coupled to matter 
with positive coupling $F>0$ investigated in Refs.~\cite{Chen, Morisaki} 
has neither ghost nor Laplacian instabilities either.

\begin{figure}[h]
\begin{center}
\includegraphics[height=3.2in,width=3.4in]{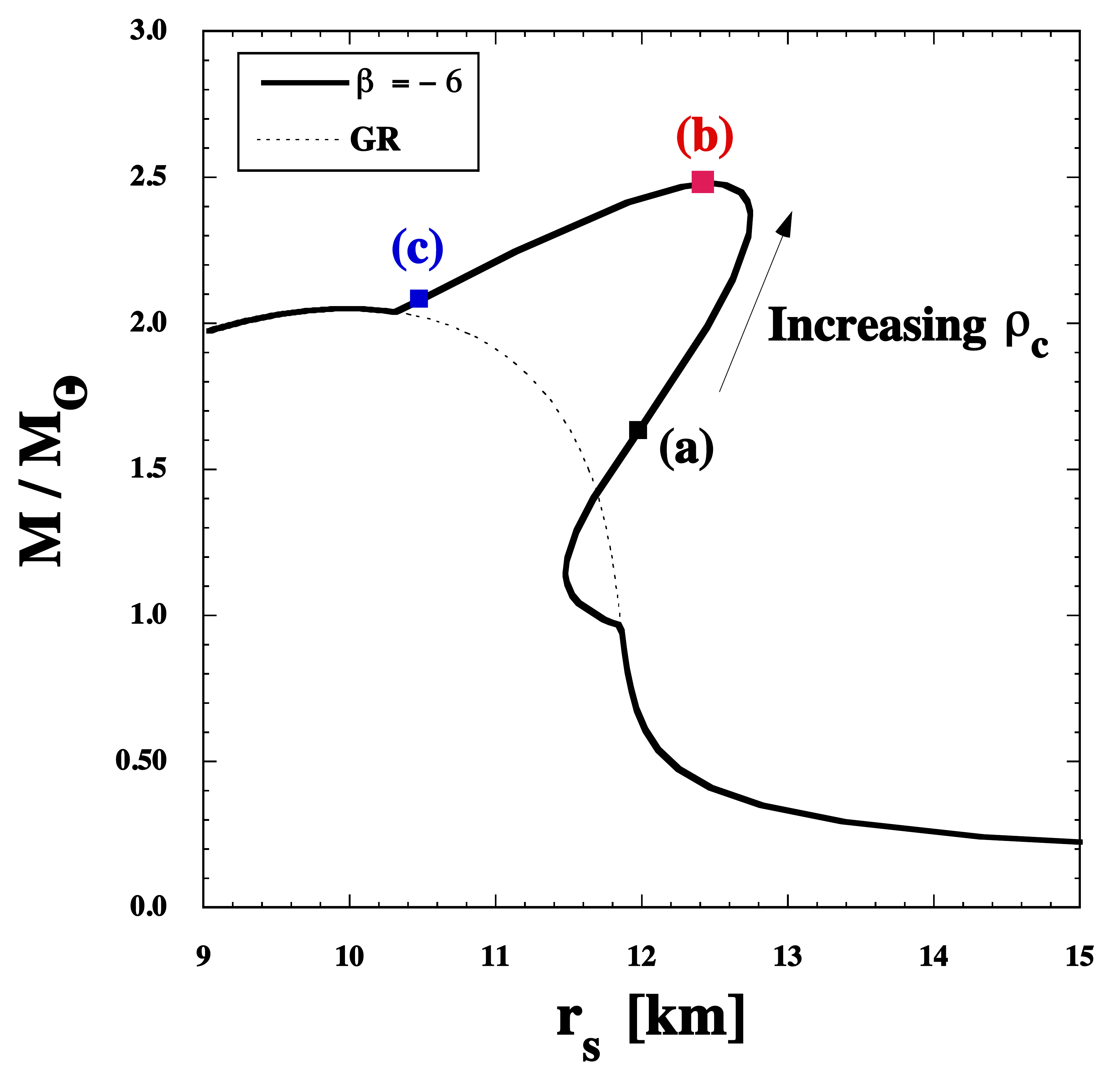}
\includegraphics[height=3.2in,width=3.5in]{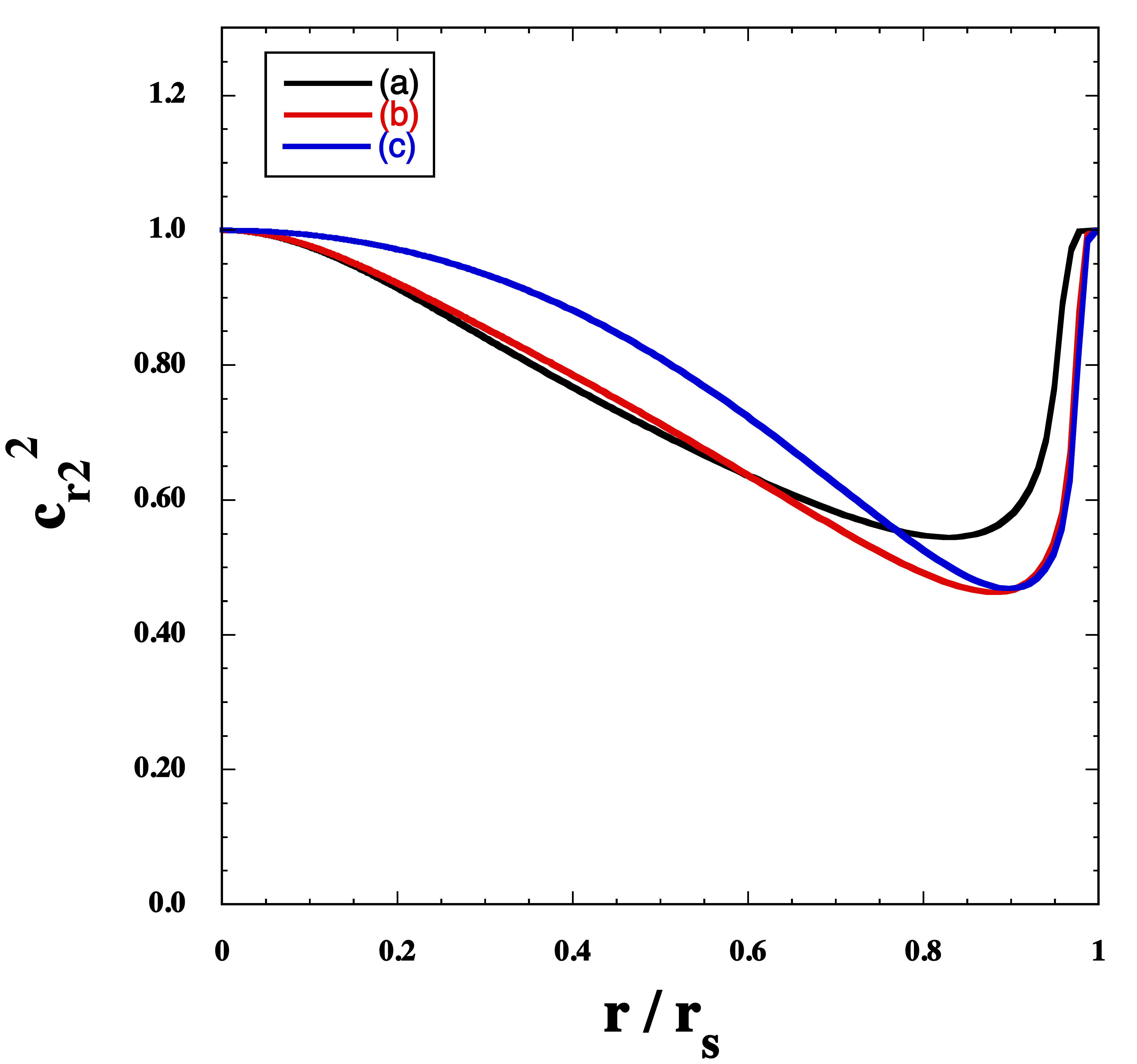}
\end{center}
\caption{\label{fig1} 
(Left) 
The mass $M$ of NS normalized by the solar mass $M_{\odot}$ 
versus the star radius $r_s$ in theories of spontaneous scalarization. 
We choose the Damour and Esposito-Farese nonminimal coupling
(\ref{Fnon}) with $\beta=-6$. 
We consider the SLy equation of state inside the star. 
For increasing central density $\rho_c$, the mass-radius 
relation moves in the direction shown as an arrow.
The mass-radius relation in GR is plotted
as a thin dashed line.
(Right)
The radial propagation speed squared $c_{r2}^2$ for $l=2$ 
versus the distance 
$r$ normalized by the star radius $r_s$. 
The plots represented as (a), (b), (c) correspond 
to the cases shown as the same labels in the left panel. 
Spontaneous scalarization occurs in all these three cases with 
the subluminal values of $c_{r2}^2$ inside the star. }
\end{figure}

In the left panel of Fig.~\ref{fig1}, we plot the mass-radius relation 
of relativistic star for the nonminimal coupling (\ref{Fnon}) 
with $\beta=-6$. 
We choose the SLy equation of state inside the star, whose 
analytic representation is given in Ref.~\cite{Haensel:2004nu}. 
For each central density $\rho_c$,  
the boundary condition of $\phi$ at $r=0$ is iteratively searched 
to realize the asymptotic behavior $\phi(r) \to 0$ as $r \to \infty$. 
For the central density $4.3 \rho_0 < \rho_c  < 14.4 \rho_0$, 
where $\rho_0 
=1.6749 \times 10^{14}~{\rm g} \cdot {\rm cm}^{-3}$, 
we find that there exists the scalarized branch 
with $\phi(0) \neq 0$ besides the GR branch with $\phi(r)=0$ 
everywhere. As the central density increases, the mass-radius 
relation shifts toward the direction of an arrow depicted in Fig.~\ref{fig1}.

Three cases (a), (b), (c) shown in the left panel of Fig.~\ref{fig1} 
are in the region where spontaneous scalarization takes place. 
In the right panel, we plot the propagation speed squared (\ref{crf1}) 
versus $r/r_s$ for $l=2$ in three different cases (a), (b), (c). 
For $\beta=-6$, we find that the matter sound speed squared 
is subluminal ($0<c_m^2 \le 1$) in the region where 
spontaneous scalarization occurs.  
In cases (a), (b), (c) of Fig.~\ref{fig1}, we have 
$c_m^2=0.56, 0.74, 0.88$ at the center of NS, 
respectively. The superluminal propagation of $c_m^2$
arises for the high central density $\rho_c \gtrsim 18 \rho_0$, 
but this is the region in which only the GR branch is present. 
In the right panel of Fig.~\ref{fig1}, we can confirm that 
$c_{r2}^2$ is subluminal inside the star for the scalarized branch.
{}From Eq.~(\ref{crf1}) the value of $c_{r2}^2$ at $r=0$ is 
equivalent to 1. For increasing $r$ from the center,  
$c_{r2}^2$ decreases from $1$.
Around the surface of star, both $\rho$ and $P$ 
rapidly drop down toward 0, so $c_{r2}^2$ begins to increase 
toward the value 1. 
Besides cases (a), (b), (c), we numerically 
confirmed that the above subluminal property generally holds 
throughout the region in which spontaneous scalarization takes place.

{}From Eq.~(\ref{crf4}), we have $(c_{r2}^2)_{l=1}=c_m^2$ at $r=0$ 
due to the boundary condition $\phi'(0)=0$.
Provided that $0<c_m^2 \le 1$, $(c_{r2}^2)_{l=1}$ 
also remains subluminal inside the star and approaches the value 1 
toward the surface. 
Taking the limit $\rho, P \to 0$ in Eqs.~(\ref{crf1})-(\ref{crf4}), 
the propagation speed squares $c_{r2}^2$, $c_{\Omega 2}^2$, 
$\left( c_{r2}^2 \right)_{l=0}$, and $\left( c_{r2}^2 \right)_{l=1}$ 
are all equivalent to 1 outside the star. 
This fact is consistent with the observations of speed 
of GWs \cite{GW170817}.

\subsection{Brans-Dicke theories}

Let us proceed to the BD theories given by the action 
(\ref{actionBD}), i.e., 
\be
G_4(\phi)=\frac{M_{\rm pl}^2}{2}e^{-2 Q\phi/M_{\rm pl}}\,,\qquad
G_2(\phi,X)= \left( 1-6Q^2 \right)e^{-2 Q\phi/M_{\rm pl}} X-V(\phi)\,.
\ee
We are considering the coupling $Q^2>0$, i.e., the BD 
parameter in the range $\omega_{\rm BD}>-3/2$. 
Then, it follows that 
\be
\kappa =
{M_{\rm pl}^2 \over 2} e^{-4 Q \phi / M_{\rm pl}}>0\,.
\label{kappaBD}
\ee
Therefore, together with the conditions (\ref{nog2}), the absence of 
ghost and Laplacian instabilities is automatically guaranteed. 
As long as $0<c_m^2 \le 1$, 
the propagation speed squares 
(\ref{crf1})-(\ref{crf4}) are subluminal 
inside the star.

\section{Conclusions}

In this paper, we studied the stability of relativistic stars 
against odd- and even-parity perturbations
in scalar-tensor theories given by the action (\ref{action}). 
Our interest is the application to hairy NS solutions which are 
known to exist in theories of spontaneous scalarization and 
BD theories (see Sec.~\ref{theorysec}).
For this purpose, we need to properly deal with the matter 
sector as a form of the perfect fluid. 
The Schutz-Sorkin action (\ref{SM}) is suitable for describing 
the perfect fluid in any background space-time. 
In Sec.~\ref{scasec}, we derived covariant equations of motion 
in scalar-tensor theories (\ref{action}) with the matter action 
(\ref{SM}) and applied them to the spherically 
symmetric and static background.

In Sec.~\ref{persec}, we decomposed the perturbations of gravity, scalar-field, 
and perfect-fluid sectors into the odd- and even-parity modes. 
To our knowledge, this type of decomposition including the perfect fluid 
as a form of the Schutz-Sorkin action 
was not addressed in the literature.
We also defined the matter density perturbation 
$\delta \rho$ and velocity potential $v$ as the quantities related to 
the number density $n$ and four velocity $u_{\mu}=J_{\mu}/(n\sqrt{-g})$, 
respectively.
The radial direction is singled out on the spherically symmetric and static 
background, in which case the velocity potential is associated
with the $\theta$ and $\varphi$ components of $u_{\mu}$.

In Sec.~\ref{oddsec}, we expanded the action (\ref{action}) up to quadratic 
order in odd-parity perturbations and obtained the second-order action 
of the form (\ref{Sodd}). The perfect fluid does not affect the evolution 
of GWs in the odd-parity sector. 
For the multipoles $l \geq 2$, there is one dynamical degree of freedom 
with the propagation speed equivalent to that of light. 
In this case, the ghost is absent under the condition $G_4>0$. 
For $l=1$, there is no dynamical propagation of odd-parity perturbations.

In Sec.~\ref{evensec}, we obtained the second-order action of even-parity 
perturbations with the multipoles $l \geq 2$ in the form (\ref{eaction}) 
after eliminating some nondynamical perturbations appearing in the 
Schutz-Sorkin action.
We found that there are three propagating degrees of freedom 
characterized by $\delta \rho_m$, $\psi$, and $\delta \phi$, 
where $\delta \rho_m$ and $\psi$ are given, 
respectively, by Eqs.~(\ref{delrhom}) and (\ref{psi}). 
After integrating out all the other nondynamical perturbations, 
the final second-order action reduces to the form (\ref{Ss2}) 
with (\ref{calX}). {}From the kinetic matrix ${\bm K}$, we 
showed that the ghosts are absent under the three conditions 
(\ref{nogo1}), (\ref{nogo2}), and (\ref{nogo3}). 
Since we only exploited the relations among coefficients in Eq.~(\ref{eaction}) 
applicable to full Horndeski theories, our no-ghost conditions 
are also valid for Horndeski theories by modifying 
the coefficients (\ref{evencof}) in the Appendix to those presented 
in Ref.~\cite{Kobayashi:2014wsa}.

The matter perturbation $\delta \rho_m$ associated with the 
even-parity sector does not propagate along 
the radial direction by reflecting the property of velocity potential 
mentioned above. In scalar-tensor theories given by the action 
(\ref{action}), the other propagation speed squares are 
given by Eq.~(\ref{crgee}). If $G_{2,XX}=0$, which is the case for 
theories of scalarized relativistic stars discussed in Sec.~\ref{theorysec}, 
the radial speeds of propagation associated with the perturbations 
$\psi$ and $\delta \phi$ reduce, respectively, 
to Eqs.~(\ref{crl>2}) and (\ref{crl>2b}). 
For $c_m^2 \neq 1$, the speed $c_{r2}$ of scalar GWs 
is different from 1 inside relativistic stars. 
In the limit $l \gg 1$, we also derived the three propagation speeds 
along the angular direction as Eqs.~(\ref{cOl>2a})-(\ref{cOl>2b}). 
Again, the perfect fluid affects the angular propagation of scalar GWs.

For the monopole mode ($l=0$) in the even-parity sector, we showed 
that there are two dynamical perturbations $v$ and $\delta \phi$ 
with the reduced action of the form (\ref{Lre}). 
In this case, the no-ghost conditions are given by Eqs.~(\ref{nogo1l=0}) 
and (\ref{nogi2l=0}) with the two radial propagation speeds 
(\ref{crl=0a}) and (\ref{crl=0}), so that the latter propagation of 
scalar-field perturbation $\delta \phi$ is modified by the
presence of perfect fluid. For the dipole mode ($l=1$), 
the dynamical perturbations correspond to $\delta \rho_m$ 
and $\psi$ defined by Eq.~(\ref{psidef2}). 
In this case, the no-ghost conditions correspond to the $l \to 1$ 
limit of those derived for $l \ge 2$. 
However, the speed of scalar GWs cannot be recovered 
in the same limit, so it gives an additional stability condition 
to those obtained for $l \ge 2$. 

In Sec.~\ref{staconsec}, we summarized the stability conditions 
for the absence of ghost and Laplacian instabilities and applied them 
to the theories with $G_{2,XX}=0$. 
Provided that the ghost is absent in the odd-parity sector ($G_4>0$) 
and that the perfect fluid satisfies the properties $\rho+P>0$ and 
$c_m^2>0$, the sign of $\kappa$ defined by Eq.~(\ref{def:kappa})
is crucial for the stability of even-parity perturbations. 
In theories of spontaneous scalarization and BD theories with $\omega_{\rm BD}>-3/2$, 
we showed that $\kappa$ is positive, under which there are neither ghosts nor
Laplacian instabilities. Moreover, as long as $0<c_m^2 \le 1$, 
the propagation speeds are subluminal inside the star. 
Indeed, we confirmed this property in theories of spontaneous scalarization 
with the nonminimal coupling taken by Damour and Esposito-Farese 
by numerically computing the radial propagation speed 
squared $c_{r2}^2$ inside the NS. 
In such theories, the propagation speeds of GWs in both 
odd- and even-parity sectors are 
equivalent to that of light outside the star, so they are 
consistent with observations of the GW170817 event.

We have thus shown that hairy relativistic stars in scalar-tensor theories 
given by the action (\ref{Jaction}) are stable against odd- and even-parity 
perturbations under mild conditions.
The next step is to probe the signature of scalar hairs from observations.
In addition to the oscillation of scalar GWs discussed in Ref.~\cite{Sotani:2005qx}, 
the tidal deformations of NS 
binaries \cite{Flanagan:2007ix,Damour:2009vw,Binnington:2009bb,Hinderer:2009ca} 
may allow one to distinguish between NSs in scalar-tensor theories and in GR.
Our general formulation of perturbations around relativistic stars will provide 
a useful framework for dealing with such problems.

\section*{Acknowledgements}

R.~Kase is supported by the Grant-in-Aid for Young Scientists B 
of the JSPS No.\,17K14297. 
ST is supported by the Grant-in-Aid for Scientific Research Fund of the JSPS No.\,19K03854.

\appendix

\renewcommand{\theequation}{A.\arabic{equation}}
\setcounter{equation}{0}

\section*{Appendix: Coefficients in the second-order action of 
even-parity perturbation}
\label{app:coefficients}

In this appendix, we summarize the coefficients of the reduced action \eqref{eaction}. 
They are given by
\ba
a_1 &=& r^2\sqrt{fh}\,G_{4,\phi}\,,\nonumber \\
a_2 &=& \frac{r}{2}\sqrt{\frac{f}{h}} \left[ hr \phi' \left( 
G_{2,X}+4 G_{4,\phi \phi} \right)+ 
G_{4,\phi} (r h'+4h) \right]= \sqrt{fh} \left( \frac{a_1}{\sqrt{fh}} \right)'
-\left( \frac{\phi''}{\phi'}-\frac{f'}{2f}  \right) a_1
+\frac{r}{\phi'} \left( \frac{f'}{f}-\frac{h'}{h} \right)a_4
+\frac{r^2}{\phi'}f_1, \nonumber \\
a_3&=& -\frac{r\sqrt{fh}}{2} \left( r\phi' G_{4,\phi}
+2G_4 \right)=-\frac{\phi'}{2}a_1-r a_4\,,\qquad
a_4= \sqrt{fh}\,G_4\,,\nonumber \\
a_5 &=& \frac{1}{2} \sqrt{\frac{f}{h}} \left[ 2G_{4,\phi} (h-1)
+2r \left( h' G_{4,\phi}+2h \phi' G_{4,\phi \phi} \right)
+r^2\{ 2(\phi'^2 G_{4,\phi \phi \phi}+\phi'' G_{4,\phi \phi})h
+h' \phi' G_{4, \phi \phi} -G_{2, \phi} \}  \right] \nonumber \\
&=& a_2' -a_1''\,,
\nonumber \\
a_6 &=& -\sqrt{\frac{f}{h}} G_{4, \phi}
=-\frac{1}{2r \phi'} \sqrt{\frac{f}{h}}
\left[ r \left( \frac{2a_4}{\sqrt{fh}} \right)'+\frac{2a_4}{\sqrt{fh}}
-{\cal F} \right]\,,\nonumber \\
a_7 &=& 
-\frac{1}{4}\sqrt{\frac{f}{h}}  \left[ 4h G_4
+4r (h' G_4+2h \phi' G_{4,\phi})+r^2 \{ h\phi'^2 (G_{2,X}+4G_{4,\phi \phi})
+2(2h\phi'' +h' \phi')G_{4,\phi} \}\right]=a_3'-\frac{r^2}{2}f_1, \nonumber \\
a_8 &=& -\sqrt{\frac{f}{h}}\frac{G_4}{2}=-\frac{a_4}{2h}\,,
\nonumber \\
a_9 &=& \frac{1}{2r} \sqrt{\frac{f}{h}} 
\left[ (r h'+2h) G_4+ 2h r\phi' G_{4,\phi} \right]
=a_4'+\left( \frac{1}{r}-\frac{f'}{2f} \right)a_4\,,\qquad
a_{10} = \frac{r^2}{2} \sqrt{\frac{f}{h}} \,, \nonumber\\
b_{1} &=& \sqrt{\frac{h}{f}}\frac{G_4}{2}
=\frac{a_4}{2f}\,,\qquad
b_{2} = -2r^2 \sqrt{\frac{h}{f}}\,G_{4,\phi} 
=-\frac{2}{f}a_1\,,\nonumber \\
b_{3} &=& -r^2 \frac{\sqrt{h}}{f^{3/2}} \left[ \phi' f 
\left( G_{2,X}+2G_{4,\phi \phi} \right)-f' G_{4,\phi} \right]
=\frac{2}{f} \left( a_1'-a_2 \right)\,,\qquad
b_{4} = r\sqrt{\frac{h}{f}} \left( r \phi' G_{4,\phi}
+2G_4 \right)=-\frac{2}{f}a_3\,,\nonumber \\
b_{5} &=&  -\sqrt{\frac{h}{f}}G_4=-2b_1\,,\nonumber \\
c_{1} &=& -\frac{r^2}{\sqrt{fh}} G_{4,\phi}
=-\frac{a_1}{fh}\,, \qquad 
c_{2}=-\frac{r}{2}\sqrt{\frac{h}{f}} 
\left[ r\phi' f (h \phi'^2 G_{2,XX}-G_{2,X})+(r f'+4f) G_{4,\phi} \right]\,,
\nonumber \\
c_{3} &=& \frac{1}{2\sqrt{fh}} \left[ fh (r^2 \phi'^2 G_{2,\phi X} 
-4r \phi' G_{4, \phi \phi}-2G_{4,\phi})+fr^2 G_{2,\phi}+2 f G_{4,\phi}
-hr f' (r \phi' G_{4,\phi \phi}+2G_{4,\phi}) \right]\,,
\nonumber \\
c_{4} &=& \sqrt{\frac{f}{h}}G_{4, \phi}\,,\qquad 
c_{5}= -\frac{1}{2r} \sqrt{\frac{h}{f}} \left[ 2fr \phi' G_{4, \phi}
+G_4(rf'+2f) \right]\,,\qquad 
\tilde{c}_{5}=-\frac{r^2(\rho+P)}{2\sqrt{h}}\,,\nonumber \\
c_{6} &=& \frac{1}{8}\sqrt{\frac{h}{f}} \left[ 4 f G_4
+4r(f' G_4+2f \phi' G_{4,\phi})+r^2 \phi' 
(fh \phi'^3 G_{2,XX}-f \phi' G_{2,X}+2f' G_{4,\phi}) \right]\,,
\nonumber \\
d_{1} &=& \sqrt{\frac{h}{f}}\frac{G_4}{2}
=\frac{a_4}{2f}\,,\qquad 
d_{2}=2 \sqrt{fh}\,G_{4,\phi}=2h c_4\,,\qquad
d_{3}=\frac{\sqrt{fh}}{r} \left[ r \phi' \left( G_{2,X}+2G_{4,\phi \phi} 
\right)-2G_{4,\phi} \right]\,,\qquad 
d_{4}=\frac{\sqrt{fh}}{r^2}G_4,
\nonumber \\
e_{1} &=& \frac{r^2}{2\sqrt{fh}}G_{2,X}
=\frac{1}{\phi' f h} \left[ \left( \frac{f'}{f}+\frac{h'}{2h} \right) a_1
+a_2-2 a_1' -2 rh a_6 \right]\,,\qquad 
e_2=-\frac{r^2\sqrt{fh}}{2} \left( G_{2,X}-h \phi'^2 G_{2,XX} \right)\,,
\nonumber \\
e_3 &=& \frac{r^2}{2} \sqrt{\frac{f}{h}} \frac{\partial {\cal E}_{\phi}}
{\partial \phi}\,,\qquad
e_4 = -\frac{1}{2} \sqrt{\frac{f}{h}} G_{2,X}\,,
\nonumber \\
f_{1} &=& -\frac{\rho+P}{2}\sqrt{\frac{f}{h}}\,,\qquad
f_{2}=-\frac{c_m^2 r^2}{2(\rho+P)} \sqrt{\frac{f}{h}}\,,\qquad
f_{3}=-\frac{r^2}{\sqrt{h}}\,,
\label{evencof}
\ea
where ${\cal F}$ and ${\cal E}_{\phi}$ are defined by
Eqs.~(\ref{nogoodd}) and (\ref{back5}) respectively.


\end{document}